%% file: manuscript_pilot.tex
\documentclass[aps,pre,twocolumn,10pt,superscriptaddress,showpacs,]{revtex4-1}

\pdfoutput=1

\usepackage[pdftex]{color,graphicx,hyperref}
\usepackage{amsmath,amsfonts,amssymb}
\usepackage[utf8]{inputenc}
\usepackage{algorithm}
\usepackage{algpseudocode}
\usepackage{tikz}
\usepackage{setspace}

\begin{document}
  
  \input{shorthand}

  \input{manuscript_preamble} 

  \input{manuscript_v4}
  
  \bibliographystyle{naturemag}
  
  \bibliography{Bibliography.bib}
  
  \onecolumngrid
  
  \clearpage
  
  \twocolumngrid
  
  \input{supplementary_preamble}
  
  \input{supplementary_v3}

\end{document}

%% file: shorthand.tex
\def\kj{k_{j}}
\def\mj{m_{j}}
\def\ki{k_{i}}
\def\mi{m_{i}}
\def\vk{\mathbf{k}}
\def\vm{\mathbf{m}}
\def\vw{\mathbf{w}}
\def\zv{\mathbf{z}}
\def\vk{\mathbf{k}}
\def\vm{\mathbf{m}}
\def\vz{\mathbf{z}}
\def\vw{\mathbf{w}}
\def\vkvw{\mathbf{k} \cdot \mathbf{w}}
\def\vmvw{\mathbf{m} \cdot \mathbf{w}}
\def\vzvw{\mathbf{z} \cdot \mathbf{w}}
\def\vzero{\mathbf{0}}
\def\qk{q_{k}}
\def\qm{q_{m}}
\def\ej{\mathbf{e}_j}
\def\ei{\mathbf{e}_i}
\def\pk{p_{k}}
\def\pw{p_{w}}
\def\pvk{p_{\vk}}
\def\zi{z_{i}}
\def\zj{z_{j}}
\def\Fkm{F_{\vk, \vm}}
\def\Rkm{R_{\vk, \vm}}
\def\fkm{f(\vk, \vm)}
\def\fkei{f(\vk, \ei)}
\def\wj{w_{j}}

\def\Cvkvm{C_{\vk, \vm}}
\def\Cvk{C_{\vk}}
\def\Ck{C_{k}}

\def\Svkvm{S_{\vk, \vm}}
\def\Svk{S_{\vk}}
\def\Sk{S_{k}}
\def\skm{s_{\vk, \vm}}
\def\skmej{s_{\vk, \vm - \ej}}
\def\skmei{s_{\vk, \vm - \ei}}
\def\dskm{\dot{s}_{\vk, \vm}}
\def\vs{\mathbf{s}}

\def\Ivkvm{I_{\vk, \vm}}
\def\Ivk{I_{\vk}}
\def\Ik{I_{k}}
\def\ikm{i_{\vk, \vm}}
\def\ikmej{i_{\vk, \vm - \ej}}
\def\ikmei{i_{\vk, \vm - \ei}}
\def\dikm{\dot{i}_{\vk, \vm}}
\def\vi{\mathbf{i}}
\def\ii{i_{i}}
\def\ij{i_{j}}

\def\vlambda{\boldsymbol{\lambda}}

\def\nuj{\nu_j}
\def\nui{\nu_i}
\def\nul{\nu_l}
\def\vmu{\boldsymbol{\mu}}
\def\vnu{\boldsymbol{\nu}}
\def\dvnu{\dot{\boldsymbol{\nu}}}
\def\dnuj{\dot{\nu}_j}
\def\dnui{\dot{\nu}_i}
\def\drho{\dot{\rho}}
\def\sumvk{\sum_{\vk}}
\def\sumvm{\sum_{\vm}}
\def\sumvkvm{\sum_{\vk, \vm}}
\def\sumj{\sum_{j=1}^n}
\def\sumi{\sum_{i=1}^n}
\def\sumLess{\sum_{S|_{F = p}}}
\def\sumMore{\sum_{S|_{F = 1}}}
\def\prodj{\prod_{j=1}^n}
\def\prodi{\prod_{i=1}^n}
\def\prodinj{\prod_{i \neq j}^n}
\def\prodlnji{\prod_{l \neq j, i}^n}

\def\betasj{\beta^s_j}
\def\betasi{\beta^s_i}
\def\betaij{\beta^i_j}
\def\betaii{\beta^i_i}
\def\vbeta{\boldsymbol{\beta}}
\def\gammasj{\gamma^s_j}
\def\gammasi{\gamma^s_i}
\def\gammaij{\gamma^i_j}
\def\gammaii{\gamma^i_i}
\def\vgamma{\boldsymbol{\gamma}}

\def\Ej{E_{j}}
\def\muj{\mu_{j}}
\def\mutau{\mu_{\tau}}
\def\sigmatau{\sigma_{\tau}}
\def\psitau{\psi(\tau)}
\def\vzeta{\boldsymbol{\zeta}}
\def\vpsihat{\boldsymbol{\hat{\psi}}}
\def\vPsihat{\boldsymbol{\hat{\Psi}}}
\def\vEhat{\boldsymbol{\hat{E}}}
\def\vmuhat{\boldsymbol{\hat{\mu}}}
\def\vnuhat{\boldsymbol{\hat{\nu}}}

\def\Bkmj{B_{k_j, m_j}}
\def\Bkmi{B_{k_i, m_i}}
\def\Bkml{B_{k_l, m_l}}
\def\Bkk1{B_{k, k_1}}
\def\Bkomj{B_{k_j - 1, m_j}}
\def\Bkmoj{B_{k_j, m_j - 1}}
\def\dBkomj{\dot{B}_{k_j - 1, m_j}}
\def\dBkmi{\dot{B}_{k_i, m_i}}

\def\gam{\delta}
\def\Tr{Tr}
\def\tauavg{\langle \tau \rangle}
\def\taumin{\tau_{0}}
\def\taumax{\tau_{1}}

\def\kmclass{(k,m)}
\def\vkvm{\textbf{k}, \textbf{m}}
\def\vkvmclass{(\textbf{k},\textbf{m})}
\def\phiz{(\phi, z)}

\def\B{Barab{\'a}si }
\def\A{Albert }
\def\BA{Barab{\'a}si-Albert }
\def\WS{Watts-Strogatz }
\def\E{Erd{\H o}s }
\def\R{R{\'e}nyi }
\def\ER{Erd{\H o}s-{R\'e}nyi }
\def\RK{Runge-Kutta }
\def\MC{Monte Carlo }

\def\ccal{\mathcal{C}}
\def\scal{\mathcal{S}}
\def\ical{\mathcal{I}}

%% file: manuscript_preamble.tex
\newcommand{\eref}[1]{Eq.~(\ref{#1})}
\newcommand{\erefs}[2]{Eqs.~(\ref{#1}) and (\ref{#2})}
\newcommand{\sref}[1]{Section~\ref{#1}}
\newcommand{\fref}[1]{Fig.~\ref{#1}}
\newcommand{\frefs}[2]{Figs.~(\ref{#1}) and (\ref{#2})}
\newcommand{\tref}[1]{Table~\ref{#1}}

\newcommand{\SNmaster}{Supplementary Note 1} 
\newcommand{\SNrandomwalk}{Supplementary Note 2} 
\newcommand{\SNedgestate}{Supplementary Note 3} 
\newcommand{\SNmontecarlo}{Supplementary Note 4} 
\newcommand{\SNlaplace}{Supplementary Note 5} 
\newcommand{\SNdiffusion}{Supplementary Note 6} 
\newcommand{\SNmeanfield}{Supplementary Note 7} 
\newcommand{\SNskewness}{Supplementary Note 8} 
\newcommand{\SNprobability}{Supplementary Note 9} 

%% file: manuscript_v4.tex
\author{Samuel Unicomb}
\email{samuel.unicomb@gmail.com}
\affiliation{Universit\'{e} de Lyon, ENS de Lyon, INRIA, CNRS, UMR 5668, IXXI, 69364 Lyon, France}
\author{Gerardo I\~{n}iguez} 
\affiliation{Department of Network and Data Science, Central European University, A-1100 Vienna, Austria}
\affiliation{Department of Computer Science, Aalto University School of Science, FI-00076 Aalto, Finland}
\affiliation{IIMAS, Universidad Nacional Auton{\'o}ma de M{\'e}xico, 01000 Ciudad de M{\'e}xico, Mexico}
\author{James P. Gleeson}
\affiliation{MACSI and Insight Centre for Data Analytics, University of Limerick, Limerick V94 T9PX, Ireland}
\author{M\'{a}rton Karsai}
\affiliation{Department of Network and Data Science, Central European University, A-1100 Vienna, Austria}
\affiliation{Universit\'{e} de Lyon, ENS de Lyon, INRIA, CNRS, UMR 5668, IXXI, 69364 Lyon, France}

\title{Dynamics of cascades on burstiness-controlled temporal networks}

\begin{abstract}
  Burstiness, the tendency of interaction events to be heterogeneously distributed in time, is critical to information diffusion in physical and social systems. However, an analytical framework capturing the effect of burstiness on generic dynamics is lacking. We develop a master equation formalism to study cascades on temporal networks with burstiness modelled by renewal processes. Supported by numerical and data-driven simulations, we describe the interplay between heterogeneous temporal interactions and models of threshold-driven and epidemic spreading. We find that increasing interevent time variance can both accelerate and decelerate spreading for threshold models, but can only decelerate epidemic spreading. When accounting for the skewness of different interevent time distributions, spreading times collapse onto a universal curve. 
  Our framework uncovers a deep yet subtle connection between generic diffusion mechanisms and underlying temporal network structures that impacts on  a broad class of networked phenomena, from spin interactions to epidemic contagion and language dynamics.
\end{abstract}

\maketitle

Temporal networks provide a representation of real-world complex systems where interactions between components vary in time~\cite{holme2012temporal, masuda2016guide, holme2015modern}. Although they were initially modelled as Poisson processes, where independent events are homogeneously distributed in time, real-world  network interactions have been found to be heterogeneously distributed and to exhibit temporal correlations~\cite{barabasi2005origin, goh2008burstiness,karsai2012universal}. In particular, interaction events in real systems concentrate within short periods of intense activity followed by long intervals of inactivity, an effect known as burstiness. Bursty dynamics appear in diverse physical phenomena including earthquakes~\cite{davidsen2013earthquake} and solar flares~\cite{deArcangelis2006universality}, biological processes like neuron firing~\cite{turnbull2005string}, and even the dynamics of human social interaction~\cite{karsai2018bursty,goh2008burstiness}.

Burstiness in temporal interactions has profound implications for the diffusion of information over temporal networks, as demonstrated in a growing number of works~\cite{karsai2011small, lambiotte2013burstiness, jo2014analytically, horvath2014spreading, williams2019effects, vazquez2007impact, mancastroppa2019burstiness}. This is true in the case of epidemic processes, often referred to as simple contagion, where the probability of infection of an uninfected node depends linearly on the number of exposures, i.e., temporal interactions with infected neighbours in the network~\cite{pastorsatorras2015epidemic}. Epidemic models successfully describe the spread of biological disease~\cite{vespignani2020modelling}, and have been shown to critically depend on burstiness and other patterns of temporal interactions~\cite{starnini2017equivalence,lambiotte2013burstiness,liu2014controlling,masuda2017temporal,masuda2020small}. Epidemic spreading over temporal networks appears to be slowed due to burstiness in some cases~\cite{karsai2011small,miritello2011dynamical,hiraoka2018correlated,min2011spreading}, while accelerated in others~\cite{rocha2011simulated}. Threshold mechanisms provide another class of phenomena where bursty temporal networks play a crucial role. Threshold dynamics, also known as complex contagion, are used to model the spread of information where infection requires the reinforced influence of at least a certain fraction of neighbours in the network~\cite{granovetter1978threshold}. Threshold driven dynamics over static networks have been extensively studied both empirically~\cite{karsai2016local} and theoretically~\cite{watts2002simple,gleeson2008cascades,karsai2016local,unicomb2018threshold,unicomb2019reentrant}, but analysis of their behaviour on temporal networks has been limited to a small number of empirical studies~\cite{karimi2013Athreshold, karimi2013Btemporal, takaguchi2013bursty, backlund2014effects}. Here we propose an analytical framework to systematically describe the relationship between the diffusion of information and bursty temporal interactions, thus providing the theoretical foundation necessary to uncover the role of burstiness in generic diffusion processes, including simple and complex contagion models of physical, biological and social phenomena.

We incorporate the most widely documented features of temporal interactions into a framework of binary state dynamics and benchmark its behaviour with standard models of threshold driven and epidemic spreading. Although stochastic bursty interactions are likely emergent phenomena~\cite{barabasi2005origin, vazquez2006modeling}, their dynamics are well approximated by renewal processes~\cite{whitt1982approximating}. Temporal heterogeneity in network interactions can then be characterised by the variability in interevent times $\tau$ (the time between consecutive events on a given edge), parameterised by the interevent time distribution $\psi (\tau)$, while other features of the temporal network are considered maximally random. Renewal processes represent the simplest model of bursty, non-Markovian dynamics, and a departure from the memoryless assumption implicit in Poisson processes. Nevertheless, we are able to show that such a system can be accurately captured by a master equation formalism, which is essentially memoryless, implying the existence of a purely Markovian system with almost identical behaviour. We show both analytically and numerically that bursty temporal interactions give rise to a percolation transition in the connectivity of the temporal network, separating phases of slow and rapid dynamics for both  epidemic and threshold models of information diffusion. We find that diffusion dynamics are sensitive to the choice of interevent time distribution, particularly in regard to its skewness, and we demonstrate a data collapse across distributions when controlling for this effect.


\textit{Temporal network model.} To model a temporal network, we consider an undirected, unweighted static network of $N$ nodes as the underlying structure, which acts as a skeleton on top of which temporal interactions take place. The degree of a node (how many neighbours it has) takes discrete values $k = 0, \ldots, N - 1$ from a degree distribution $\pk$. Pairwise temporal interactions, or {\it events}, occur independently at random on each static edge via a renewal process 
with interevent time distribution $\psi (\tau)$. Time is continuous and events are instantaneous, while consecutive interevent times are uncorrelated.
We also assume that the renewal process is stationary (for further details see Methods and \SNmontecarlo{}). By using a static underlying network, we assume the time scales of edge formation and node addition or removal are far longer and thus negligible relative to the time scale of event dynamics over existing edges.

In its simplest form, information diffusion is a binary-state process where each node occupies one of two mutually exclusive states, which we term \textit{uninfected} and \textit{infected}. The probability of a node changing state is a function of the state of its neighbours, as well as the strength of their interactions. Interaction strength, also referred to as mutual \textit{influence}, is a non-negative scalar that we consider to be a function of the elapsed renewal process time series. We desire that the mean of the emergent distribution of interaction strengths be stationary, and invariant to the underlying burstiness of the system. This is achieved when the contribution of a single event to interaction strength \textit{(i)} goes to zero as the event ages, and \textit{(ii)} is additive, meaning a spike in edge activity leads to a spike in the interaction strength between neighbours. Under these assumptions, the simplest such coupling is a step function, i.e., the contribution of an event to interaction strength is constant for a duration $\eta$, after which it goes to zero. As such we define the interaction strength $w_j$ of an edge at time $t$ (or state $j$ for short), as the number $j$ of events having occurred in the preceding time window of width $\eta$.

It follows that the local configuration of a node is determined by the number $k_j$ of its neighbours connected via edges in state $j$, with the degree $k$ of the node related to its $k_j$ values by $k = \sum_j k_j$ at any time $t$. We introduce $m_j$ as the number of infected neighbours of a node connected via edges in state $j$. Consequently, $0 \leq m_j \leq k_j$ with $m = \sum_j m_j$ the total number of infected neighbours. For each node, we store $k_j$ and $m_j$ for all $j$ in vectors $\vk$ and $\vm$, providing a description of edge and node states in the local neighbourhood of a node. Nodes in class $\vkvmclass$ become infected at a rate $\Fkm$, and are statistically identical in this sense. We also store the interaction strength $w_j = j$ in the vector $\mathbf{w}$ for all $j$. The dynamics of the influence received by a node is thus fully determined by $\vkvmclass$ and $\vw$.

\textit{Models of information diffusion.} To examine the  effect of temporal interactions on information diffusion, we explore three widely known models of transmission. We consider both relative (RT) and absolute (AT) variants of a threshold mechanism~\cite{watts2002simple, granovetter1978threshold,centola2007complex}, as well as the Susceptible-Infected (SI) model of epidemic spreading~\cite{valdano2015analytical} (see \tref{tab:tab1} for details). All models are non-recovery, meaning the uninfected state cannot be reentered, and we consider infection due to external noise at a low, but nonzero rate $p$.

\begin{table}[t]
  \caption{Transmission rate $\Fkm$ for nodes in configuration $(\textbf{k}, \textbf{m})$ with interaction strength $\mathbf{w}$ and infection rate $p$ due to external noise. In complex contagion models with relative (RT) and absolute (AT) thresholds, infection is regulated by parameters $\phi$ and $M_\phi$, respectively. In the Susceptible-Infected (SI) model, infection is determined by the rate $\lambda$.
  \label{tab:tab1}}
	\begin{ruledtabular}
		\begin{tabular}{c c c}
        \multicolumn{2}{c}{threshold} & \\
	    \hspace*{-3mm}relative & \hspace*{-5mm}absolute & \hspace*{-3mm}SI \\
	  	\hline 
	    $\begin{cases} 1,\enspace  \mathbf{m} \cdot \mathbf{w} \geq \phi \mathbf{k} \cdot \mathbf{w} \\ p, \enspace \text{otherwise}\end{cases}$
	    &
	    \hspace*{-3mm}$\begin{cases} 1,\enspace  \mathbf{m} \cdot \mathbf{w} \geq M_{\phi} \\ p, \enspace \text{otherwise}\end{cases}$
	    &
	    \hspace*{-3mm}$\max(p,\ \textbf{m} \cdot \boldsymbol{\lambda})$\\
		\end{tabular}
	\end{ruledtabular}
\end{table}	

The study of threshold dynamics focuses on the conditions leading to {\it cascades}, or large avalanches of infections that sweep through the network. In the simplest implementation of threshold dynamics, infection occurs when the number $m$ of infected neighbours of an uninfected node exceeds a fraction $\phi$ of its degree $k$~\cite{watts2002simple, granovetter1978threshold}. Generalising this rule to the case or arbitrary interaction strength~\cite{yagan2012analysis, unicomb2018threshold}, in the RT model infection occurs when the \textit{influence} of infected neighbours, $\vm \cdot \vw$, exceeds a fraction $\phi$ of all potential influence, $\vk \cdot \vw$. The RT model captures instances of real-world diffusion where interaction between elements affect the probability of infection only in aggregate, similar to the response of individuals to new behavioural patterns or transmission in biological neural networks~\cite{gerstner2014neuronal, iyer2013influence}. When considering the RT model over temporal networks, the probability of infection may increase during bursts of interaction events with infected neighbours or, conversely, bursts of activity with uninfected neighbours may temporarily maintain a node in the uninfected state. In the AT model, influence from infected neighbours is not normalised, but compared to some absolute value $M_\phi$~\cite{centola2007complex}. In contrast to the RT model, infection is not hindered by interaction activity with uninfected neighbours, and bursts can only increase the probability of infection. In the SI model, finally, each interaction event with an infected neighbour triggers infection at a rate $\lambda$. In our framework of temporal networks, infected neighbours trigger infection via edges in state $j$ at a rate $\lambda j$. Writing $\vlambda = \lambda \vw$, the infection rate for a node with a neighbourhood of infected nodes described by $\vm$ is $\textbf{m} \cdot \vlambda$. Similar to the AT model, bursts can only increase the probability of infection in the SI model. 

\textit{Binary dynamics over temporal networks.} We extend a master equation formalism~\cite{gleeson2011high, unicomb2018threshold} to account for network temporality. We introduce the state space of all configurations $\vkvmclass$ allowed by the underlying degree distribution $p(k)$, under the condition that each edge is in one of a finite number of possible edge states (see Methods, and \SNmaster{} for lattice diagrams of this space). We introduce the state vector $\mathbf{s}(t)$ containing the probability that a randomly selected node with underlying degree $k$ is uninfected and in class $\vkvmclass$ at time $t$. The time evolution of $\mathbf{s}$ is governed by the matrix $W(\mathbf{s}, t)$, containing the transition rate $W_{ij}$ from the $i$-th to the $j$-th configuration $\vkvmclass$ at time $t$. Transitions arise from three mechanisms. First, \textit{ego} transitions, contained in the matrix $W_{ego}$, describe the loss to configuration $\vkvmclass$ due to its nodes becoming infected. This occurs at a rate $\Fkm$, as per \tref{tab:tab1}, so the diagonal terms of $W_{ego}$ are given by $-\Fkm$ and off-diagonals are zero. Second, \textit{neighbour} transitions, contained in matrix $W_{neigh}$, describe the gain or loss to configuration $\vkvmclass$ due to the infection of neighbours of nodes in this class. This transition is determined by $\beta_j dt$, the probability of an uninfected neighbour in configuration $j$ becoming infected over an interval $dt$ (see Methods for an explicit calculation). Taken together, $W_{ego}$ and $W_{neigh}$ accurately describe diffusion
dynamics over static and heterogeneously distributed edges, such as weighted and multiplex networks~\cite{unicomb2018threshold, unicomb2019reentrant}. 

Temporal networks require a third component, \textit{edge} transitions, contained in the matrix $W_{edge}$, describing the gain or loss to configuration $\vkvmclass$ due to changes in an edge's state $j$. This applies to any temporal network model that can be formulated in terms of discrete, dynamic edge states. We denote by $\mu_j dt$ and $\nu_j dt$ the probabilities that a randomly selected edge in state $j$ undergoes a positive or negative transition and enters state $j + 1$ or $j - 1$, respectively, over an interval $dt$. Combining these terms gives the master equation
\begin{equation}\label{eqn:master}
  \dfrac{d}{dt} \mathbf{s} = (W_{ego} + W_{neigh} + W_{edge})\mathbf{s} = W(\mathbf{s}, t)\mathbf{s}.
\end{equation}
Modelling temporal network dynamics amounts to solving \eref{eqn:master}, which along with the initial condition $\mathbf{s}(0)$, determine the evolution of the system.

\begin{figure}[t]
  \hspace*{-1.4mm}
  \includegraphics[scale=1]{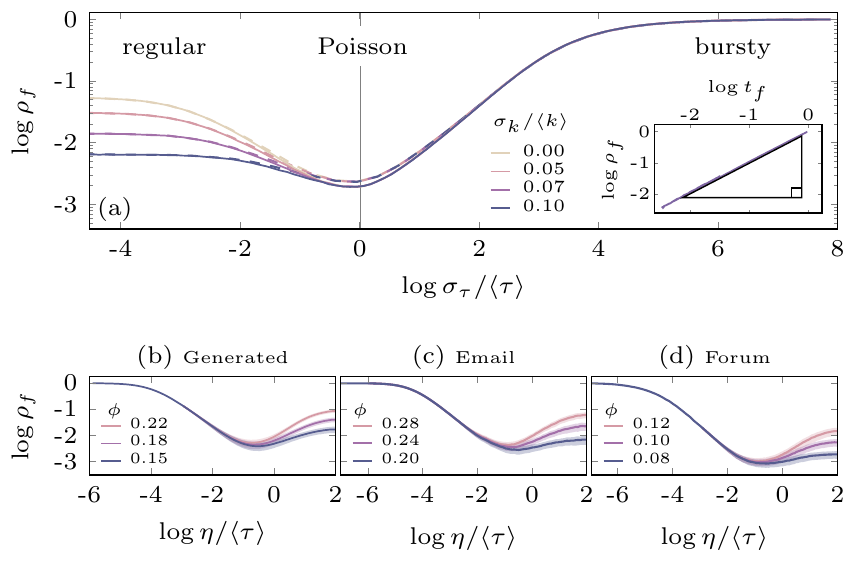}
  \caption{Normalised density of noise-induced infections $\rho_f$ as a function of interevent time standard deviation $\sigma_\tau$ and memory $\eta$. Normalised diffusion time $t_f$ produces an almost identical effect [see (a), inset]. The small $\sigma_\tau$ limit, leading to regular patterns in $\tau$, comprises the quenched limit in (a) where $\eta = 1$ and the network is effectively static. The large $\sigma_\tau$ limit produces large bursts in activity, comprising the annealed regime where the network is effectively sparsified and plays no role in information diffusion ($\rho_f = 1$).  Mirroring results are achieved by varying memory $\eta$ for fixed $\sigma_\tau$ in generated (b) and empirical (c-d) temporal networks. Analytic solution is denoted by dashed lines, and Monte Carlo results by solid lines. Generated networks have lognormal degree distribution with mean $\langle k \rangle = 7$ and standard deviation $\sigma_k = 2$. We use Weibull-distributed interevent times with mean $\langle \tau \rangle = 1$. Plot (b) uses $\sigma_\tau = 1$. For empirical data description see Methods. Node dynamics correspond to the RT model with threshold $\phi = 0.15$ and external noise $p = 2 \times 10^{-4}$. Cutoff density is $\rho_c = 0.4$. Monte Carlo
  simulations are averaged over $10^4$ realisations. Network size is $10^6$ in (a), and $5\times 10^3$ in (b).\label{fig:fig2}}
\end{figure}

To apply this formalism we derive the edge transition rates $\mu_j$ and $\nu_j$ in the case of renewal processes. We first note that microscopically, on the scale of a single edge, transitions from state $j$ to $j \pm 1$ cannot be described by a constant rate. In a renewal process, the probability of an event occurring is conditional on the time elapsed since the previous event. Therefore, this probability is history dependent, meaning edges have an effective memory and are non-Markovian by definition. Further, since it is only the previous event that is determinant, there is clearly no $j$ dependence at this scale. A renewal process may then seem at odds with a Markovian master equation [where $\mathbf{s} (t + dt)$ depends only on $\mathbf{s} (t)$, as per \eref{eqn:master}]. Macroscopically however, on the scale of large ensembles of edges, the renewal process exhibits effective $j$-dependent rates that are constant in time. We can calculate the probability $E_j$ that a randomly selected edge is in state $j$, and the probability that it transitions to state $j \pm 1$ over an interval $dt$, giving $\mu_j$ and $\nu_j$ [see Methods for explicit expressions for $j > 0$, with the $j = 0$ case of $E_j$ and $\mu_j$ comprising a special case that we define in \erefs{eqn:xiE}{eqn:ximu} below].

Since the rates $\mu_j$ and $\nu_j$ are heterogeneous in terms of $j$, they can be viewed as a signature of the model parameters $\psi(\tau)$ and $\eta$, and of the non-Markovianity inherent at the scale of a single edge. On a macroscopic scale, $\mu_j$, $\nu_j$, and $E_j$ are constant in time, meaning our system is indistinguishable from a continuous-time Markov chain model of edge state. That is, a random walk on the non-negative integers, with transition rates given by $\mu_j$ and $\nu_j$, and a stationary distribution of walkers given by $E_j$ [see \SNrandomwalk{} for an illustration of $\mu_j$ and $\nu_j$ in the case of a gamma distribution $\psi(\tau)$]. Applying the system-wide rates $\mu_j$ and $\nu_j$ at the finer-grained level of configurations $\vkvmclass$ amounts to a mean field approximation. Monte Carlo simulations (see Supplementary Fig. 8) demonstrate that the actual edge transition rates deviate slightly from $\mu_j$ and $\nu_j$ for each configuration $\vkvmclass$, even if they are exact for the network as a whole, in the limit of large $N$. The accuracy of the master equation solution provides a measure of the remarkable similarity between a renewal process, where \eref{eqn:master} is an approximation, and the biased random walk interpretation of edge state, where \eref{eqn:master} is exact. 

\textit{Burstiness and information diffusion.} We validate our analytical framework with Monte Carlo simulations of diffusion dynamics over temporal networks. Simulations use an underlying static, configuration-model network with lognormal degree distribution of mean $\langle k \rangle$ and standard deviation $\sigma_k$. We measure the time $t_c$ required to reach an arbitrary density $\rho_c$ of infected nodes, in the presence of background noise at rate $p$. We also measure $\rho_f$, the relative frequency of infections due to external noise, such that $0 < \rho_f \leq 1$, with $1 / \rho_f$ the ratio of all to noise-induced infections, measuring the catalytic effect of external noise (for a detailed description of $\rho_f$ see Methods). We normalise $t_c$ by the time taken to reach the desired density by noise only, providing $t_f$, such that $0 < t_f \leq 1$. Remarkably, $\rho_f$ and $t_f$ are almost equivalent, with a value of $\rho_f = t_f = 1$ indicating slow diffusion with complete reliance on external noise, and small $\rho_f$ and $t_f$ representing rapid diffusion with external noise producing a substantial catalytic effect. Together, they measure the extent to which the temporal network, rather than external noise, drives the diffusion of information. 

\begin{figure}[b]
  \centering
  \hspace*{-1.2mm}
  \vspace*{-1mm}
  \includegraphics[scale=1]{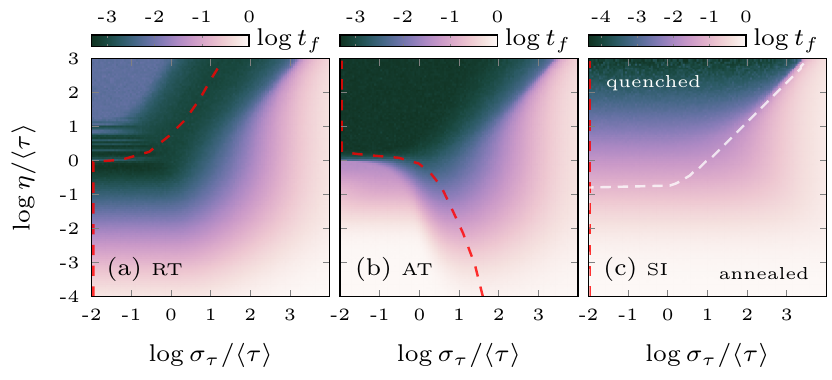}
  \caption{Monte Carlo simulation of the normalised diffusion time $t_f$ as a function of interevent time standard deviation $\sigma_\tau$ and memory $\eta$. Diffusion dynamics correspond to the RT model with $\phi = 0.15$ (a), the AT model with $M_\phi = 2$ (b), and the SI model with $\lambda = 0.02$ (c). Red dashed lines indicate the $\sigma_\tau$ value producing the minimum diffusion time $t_f$, for a given $\eta$. This demonstrates, for example at large $\eta$ in (a), that small values of burstiness maybe be accelerative with respect to the Poisson case. This is evidenced by the minimum extending beyond $\sigma_\tau = 1$. Oscillations in the line between $0 < \log \eta < 1$ in (a) are smoothed for clarity. White dashed line in (c) indicates the theoretical emergence of a giant connected component in the subgraph formed by removing $j = 0$ edges [identical but not shown in (a) and (b)]. Degree distribution is lognormal with $\langle k \rangle = 7$ and $\sigma_k = 0.5$. Results correspond to a single realisation of each $(\sigma_\tau, \eta)$ value in networks of size $N = 10^5$. The interevent time distribution $\psi(\tau)$ is Weibull with mean $\langle \tau \rangle = 1$ (see \SNdiffusion{} for corresponding $\rho_f$ values). 
  \label{fig:fig3}}
\end{figure}

We first examine the effect of varying interevent time standard deviation $\sigma_\tau$ for fixed memory $\eta = \langle \tau \rangle = 1$ [\fref{fig:fig2}(a)]. We choose a Weibull interevent time distribution $\psi(\tau)$, used widely to model behavioural bursts in both human \cite{jiang2013calling} and animal \cite{sorribes2011origin} dynamics. A Weibull distribution reduces to the exponential distribution for $\sigma_\tau = \langle \tau \rangle = 1$. Node dynamics follow the RT model for threshold $\phi = 0.15$ and background noise $p = 2 \times 10^{-4}$. Approaching the small $\sigma_\tau$ limit from above, events arrive in an increasingly regular pattern, and an increasing fraction of edges are frozen in the mean state $\eta / \langle \tau \rangle = 1$. We refer to this as the \textit{quenched} regime, whereby edges converge to a single state and the network is effectively static. In the opposing limit of large $\sigma_\tau$, burstiness means that at any given time, edge activity is concentrated among an arbitrarily small fraction of edges that undergo large spikes in activity, with the remainder in state $j = 0$. We refer to this as the \textit{annealed} regime, where the network is maximally sparse and has a vanishingly small role in information diffusion ($\rho_f$ and $t_f$ approach one).

Both quenched and annealed regimes lead to slow, noise-reliant diffusion, where the expected edge state $\eta / \langle \tau \rangle$ is preserved [\fref{fig:fig2}(a)]. For intermediate values of $\sigma_\tau$ there is a \textit{well-mixed} regime where relatively rapid diffusion is due to edge state fluctuations that are ultimately favourable to transmission. In the RT model this implies a spike of activity on an infected neighbour overcoming a node's threshold, or decreased activity on uninfected edges lowering the relative influence to be overcome. The decelerative effect of quenching is increased for narrower  underlying degree distributions, since an increasing fraction of nodes are frozen in a state unfavourable to transmission, a static network effect already reported in~\cite{watts2002simple}. 

A mirroring effect can be obtained by varying memory $\eta$ for constant $\sigma_\tau = \langle \tau \rangle = 1$ [\fref{fig:fig2}(b)]. The quenched limit is recovered for large $\eta$, as large samples of events on each edge result in edges converging to a mean state, $\eta / \langle \tau \rangle$, with an increasingly narrow distribution, due to the central limit theorem. As for the case of fixed $\eta$, quenching may be decelerative if cascades on the corresponding static network are noise dependent. For example, increasing $\phi$ can cause slower diffusion in the quenched limit [\fref{fig:fig2}(b)]. The annealed (noise-driven) regime is effectively recovered when $\eta$ is vanishingly small, meaning almost all edges are in state $j = 0$ and the role of the network in information diffusion vanishes ($\rho_f = t_f = 1$). The correspondence between $\sigma_\tau$ and $\eta$ suggests data-driven experiments that allow an indirect inference of the effects of varying $\sigma_\tau$ in real systems, an open problem in the study of information diffusion. We simulate the RT model on two empirical temporal networks and vary only the memory $\eta$, recovering qualitatively the effects observed on synthetic networks [\fref{fig:fig2}(c-d), see Methods for data description]. This suggests the accelerative and decelerative effects of burstiness may well be a feature of real-world information diffusion.

\begin{figure*}[t]
  \hspace*{-1.5mm}
  \includegraphics[scale=1]{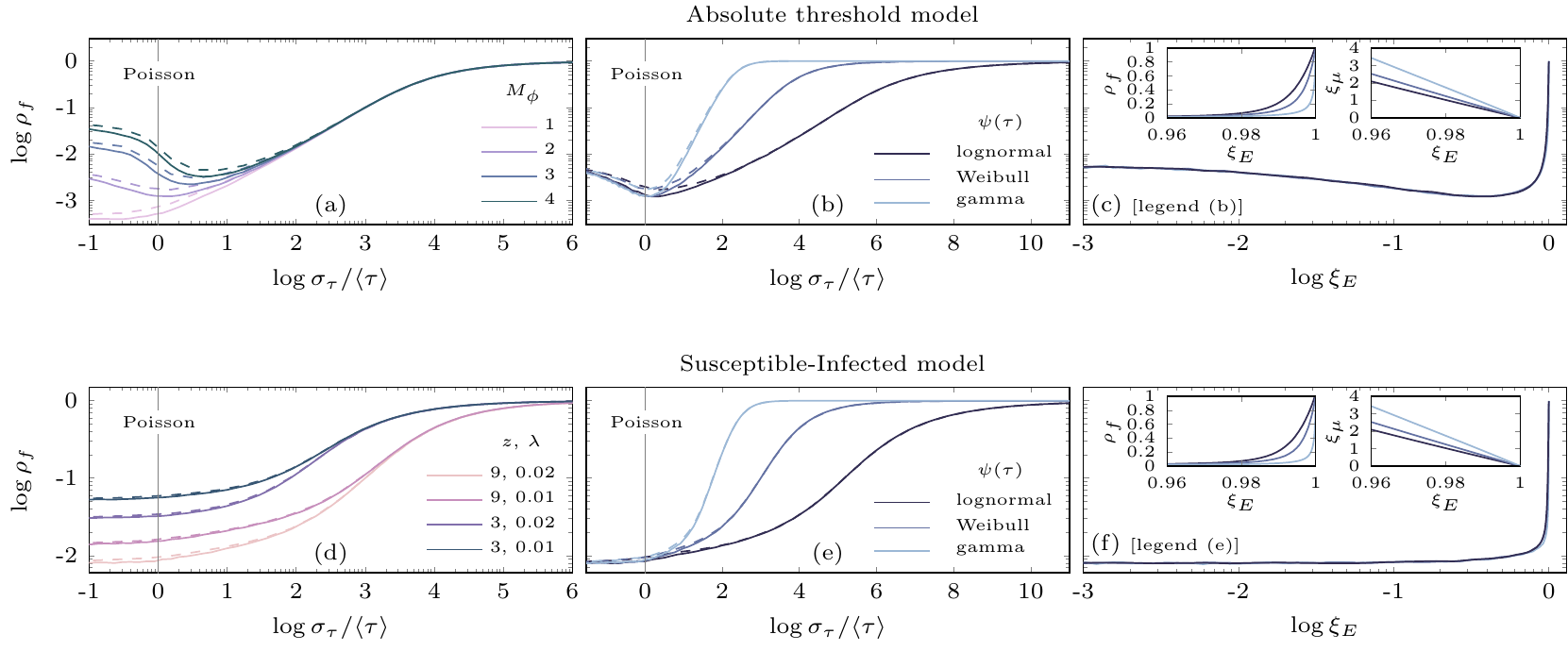}
  \caption{Relative density of noise-driven infections $\rho_f$ as a function of interevent time standard deviation $\sigma_\tau$ for the AT (a-c) and the SI (d-f) models. Analytic solution is denoted by dashed lines, and Monte Carlo results by solid lines. (a) Effect of relative threshold $M_\phi$. (b) Dependence of $\rho_f$ on choice of $\psi (\tau)$. (c) Data collapse of (b) after controlling for effective sparsity $\xi_E$. Left inset is a closeup of the main plot in linear scale, revealing differences in $\rho_f$ remain after controlling for $\xi_E$. This is explained by the differing mixing rates $\xi_\mu$ (inset right). (e-f) Similar results for the SI model. Cutoff node density is $\rho = 0.4$, with $\eta = \langle \tau \rangle = 1$.  Analytic solution is shown by dashed lines, Monte Carlo simulations by solid lines. Degree distribution is lognormal with $\langle k \rangle = 7$ and $\sigma_k = 1$. We use a threshold $M_\phi = 2$ for the AT model, and external noise $p = 2 \times 10^{-4}$. Monte Carlo simulations involve $10^4$ realisations on networks of size $N = 10^6$.\label{fig:fig4}}
\end{figure*}

We systematically explore the RT, AT, and SI models with Monte Carlo simulations over $(\sigma_\tau, \eta)$-space (\fref{fig:fig3}). The underlying degree distribution is lognormal with $\langle k \rangle = 7$ and $\sigma_k = 0.5$, and interevent times are Weibull-distributed with $\langle \tau \rangle = 1$. Our aim is to understand how the temporal connectivity evolves over $(\sigma_\tau, \eta)$-space. As previously observed, the quenched regime appears either in the small $\sigma_\tau$ limit for constant (but sufficiently large) $\eta$, or in the large $\eta$ limit for constant $\sigma_\tau$. The temporal network enters the annealed regime in two ways, either by taking the small $\eta$ limit for constant $\sigma_\tau$, or the large $\sigma_\tau$ limit for constant $\eta$. The two regimes are separated by a percolation transition, i.e., the emergence of a giant connected component in the subgraph formed by edges in state $j > 0$ [see regimes and boundary in \fref{fig:fig3}(c)]. To quantify this transition, we introduce 
\begin{equation}\label{eqn:xiE}
  \xi_E = \int_{\eta}^{\infty}\Psi (\tau)d\tau
\end{equation}
and
\begin{equation}\label{eqn:ximu}
  \xi_{\mu} = \dfrac{\int_{\eta}^{\infty} \psi (\tau)d\tau}{\int_{\eta}^{\infty} \Psi (\tau)d\tau},
\end{equation}
where $\xi_E$ equals $E_0$, the density of edges in state zero, and $\xi_\mu$ equals $\mu_0$, the probability that a randomly selected edge in state zero enters state one over an interval $dt$. We may refer to $\xi_E$ as the effective sparsification, or alternatively, the effective annealing. Here $\Psi(\tau)$ is the complementary cumulative distribution relating to $\psi (\tau)$. We denote by $q(k)$ the degree distribution obtained by randomly removing a fraction $\xi_E$ of edges in a static configuration model network with degree distribution $p(k)$, which is identical to the expected subgraph formed by removing state zero edges in the stochastic temporal network. The percolation transition for $q(k)$ can be computed analytically (see \SNedgestate{}), and despite the static assumption, provides an excellent estimate of the boundary between quenched and annealed regimes (\fref{fig:fig3}), indicating the onset of slow, noise-dependent diffusion for all diffusion dynamics considered.

Even if the outcome of information diffusion (as measured by $t_f$) is qualitatively similar across diffusion models with respect to the features of the temporal network ($\sigma_\tau$ and $\eta$), we can identify differences due to node dynamics by measuring the values of $\sigma_\tau$ that produce a minimum diffusion time for given $\eta$ (see dashed lines in \fref{fig:fig3}). In the RT model, both quenched and annealed regimes produce relatively slow diffusion. Between the two regimes, the minimum  diffusion time shifts to larger $\sigma_\tau$ for increasing memory $\eta$, and eventually exceeds $\sigma_\tau = 1$, meaning burstiness is accelerative. As $\eta$ increases, larger and larger fluctuations in $\tau$ are required to exit the quenched regime and enter the mixing phase where rapid diffusion occurs, a consequence of the central limit theorem [\fref{fig:fig3}(a)]. Increasing $\sigma_\tau$ also produces an accelerative effect in the AT model [\fref{fig:fig3}(b)]. In contrast to the RT model, burstiness is accelerative only for small values of memory $\eta$. Since in the AT model we do not normalise infectious influence by total influence, increasing $\eta$ is always favourable to transmission, and quenching never slows down diffusion [see \fref{fig:fig4}(a)]. In the SI model burstiness does not have an accelerative effect [\fref{fig:fig3}(c) and \fref{fig:fig4}(d)], since its infection rate is unbounded (as opposed to threshold models, as per \tref{tab:tab1}). The annealing effect of burstiness overwhelms the increased rate of local diffusion afforded by unbounded transmission rates, which increases proportionally to the size of the burst. As such, acceleration due to burstiness appears to be a hallmark of threshold mechanisms, whether relative or absolute.

Finally, we examine the effect of the choice of interevent time distribution $\psi(\tau)$ (\fref{fig:fig4}). We measure the noise dependence $\rho_f$ for lognormal, Weibull and gamma distributions, controlling for both $\langle \tau \rangle$ and $\sigma_\tau$. Consider first the AT model with $M_\phi = 2$ and $\eta = \langle \tau \rangle = 1$ [\fref{fig:fig4}(b)]. Here, we observe a striking dependence on $\psi (\tau)$, with the lognormal distribution leading to the most rapid diffusion, outpacing the gamma distribution in diffusion speed and relative noise dependence by up to a factor of $83$, and the Weibull distribution by up to a factor of $14$. These differences can be accounted for by comparing the rate of onset of annealing in terms of $\xi_E$ as we increase $\sigma_\tau$. The gamma distribution rapidly anneals the network, yielding the largest $\xi_E$ values of all choices of distribution, meaning the most edges in state $j = 0$. As a result, it exhibits the slowest, most noise reliant diffusion. In terms of the value of $\xi_E$ induced, the gamma is followed by the Weibull distribution, then the lognormal distribution. In fact, the lognormal requires order-of-magnitude larger $\sigma_\tau$ to produce equal values of $\xi_E$ as the Weibull and gamma distributions. By plotting $\rho_f$ against $\xi_E$ we observe the data to collapse approximately onto a single curve, revealing $\xi_E$ to be a far better predictor of dynamics than $\sigma_\tau$ [see \fref{fig:fig4}(c) in contrast to \fref{fig:fig4}(b)]. Some disagreement persists, however [\fref{fig:fig4}(c), left inset], which can be explained by noting that increased rates of mixing $\xi_\mu$ [\fref{fig:fig4}(c), right inset] ensure that the small number of active edges redistribute about the network at a greater rate, thus mediating cascades more effectively. An identical effect is observed for the SI model [\fref{fig:fig4}(e-f)].

The data collapse in \fref{fig:fig4}(c) and (f) confirms that above all it is $\xi_E$, the density of edges in state $j = 0$, that ultimately determines the diffusion dynamics in our framework. It remains to determine why the value of $\xi_E$ is so sensitive to the choice of interevent time distribution $\psi$, and in particular, what the properties are of a given distribution $\psi$ that most contribute to the value of $\xi_E$, beyond its mean and standard deviation. We have found two properties that correlate with our observations in \fref{fig:fig4}(b) and (e), at least qualitatively, as shown in Supplementary Fig. 9. They are the third raw moment, $\langle \tau^3 \rangle$, which is closely related to skewness, and differential entropy (as defined in \SNskewness{}). These measures provide a rule of thumb such that, for instance, given two distributions $\psi$ with equal mean and variance, it is the one with the greater skewness that produces the lowest $\xi_E$, and the most rapid diffusion.


\textit{Discussion}. Our study shows that generic dynamics of information diffusion are closely tied to the level of burstiness in the underlying temporal network. By considering three binary-state models of transmission, we have demonstrated that they differ in their response to burstiness only in their details. For instance, while having a purely decelerative effect on SI models, increasing burstiness at intermediate values can be accelerative for threshold models. Nevertheless, the prevailing trend is that increasing burstiness is strongly decelerative overall, with the onset of the decelerative phase heavily dependent on the choice of interevent time distribution. The key assumptions here are that the underlying network is fixed, and that due to a memory mechanism, a fraction of edges enter a non-interacting state due to long waiting times. These assumptions result in a temporal network topology that has profound implications for many dynamical processes. It is likely that structural features of the temporal network, such as the percolation transition separating slow and fast diffusion, and the data collapse observed when controlling for the effective sparsity, will also be critical for the more general class of binary-state dynamics, including not only threshold models and models of disease, but language, voter, and Ising models, among others. 

Our master equation formalism can be extended to a broad class of temporal network models. In particular, any model that can be formulated in terms of discrete, dynamic edge states is a candidate for our approach. This includes growing, decaying and adaptive networks, as well as models of rewiring. In line with our use of renewal processes, a large family of point processes have natural descriptions in terms of discrete edge states, such as cascading Poisson and Cox processes. Extensions to the Poisson process in general suggest promising applications of our approach. In particular, our treatment of non-Markovianity could be applied to other systems. That is, while a single component in a large system may be strongly non-Markovian, as was the case in our renewal process, stationary statistics may emerge at an ensemble level that act as a signature of the non-Markovianity occurring microscopically. Our biased random walk interpretation of the renewal process model shows that strikingly similar Markovian counterparts may be available for analysis.
Incidentally, biased random walk models of edge state suggest a broad class of Markovian models to which our master equation applies exactly. These may be extended, for example, to L{\'e}vy flights, and used as a probe of various complex systems where memory is critical.

\subsubsection*{Acknowledgements}

S.U. acknowledges the P{\^o}le Scientifique de Mod{\'e}lisation Num{\'e}rique from ENS Lyon for their computing support, as well as  L. Taulelle and V. Lef{\`e}vre for useful technical advice. G.I. acknowledges partial funding by the European Commission through H2020 project HumanE AI under G.A. No. 761758. J.P.G.~is supported by Science Foundation Ireland (grant numbers 16/IA/4470, 16/RC/3918, 12/RC/2289 P2 and  18/CRT/6049) with co-funding from the European Regional Development Fund. M.K. is supported by the DataRedux (ANR-19-CE46-0008) and SoSweet (ANR-15-CE38-0011) projects funded by ANR and the SoBigData++ (H2020-871042) project.

\subsubsection*{Author contributions}

S.U. derived the analytical solution, and performed all related calculations and numerical experiments. S.U., G.I., J.P.G., and M.K. designed the research and contributed to the manuscript.

\subsubsection*{Competing interests}

The authors declare no competing interests.


\appendix

\section*{M\lowercase{ethods}}

\textit{Master equation configuration space}. We provide here a minimal description of the master equation formalism, with a focus on class transition rates, with a complete description provided in \SNmaster{}. We introduce $\Cvkvm$, the set of all nodes in the network with local configuration $\vkvmclass$, such that $\vzero \leq \vm \leq \vk$.  Whereas $\Cvkvm$ is a set of nodes, we define $\Ck$ as the set of all sets $\Cvkvm$ with total degree $k$. This can be written $\Ck = \lbrace \Cvkvm \mid \sum_j\kj = k \rbrace$.  Then, we refer to the configuration space $C$ as the set of all possible sets $\Cvkvm$. Given a degree distribution $\pk$, we define
\begin{equation}
  C = \{\vkvmclass \mid k \in \text{supp}\ p(k) \enspace \text{and} \enspace \vzero \leq \vm \leq \vk\},
\end{equation}
which partitions the network at any given time. Written this way, $C$ is potentially infinite. To ensure that it be finite in numerical constructions, we assume an upper cutoff in the degree distribution $p(k)$, and the set of edge states to be of a finite size $n$. Note that $C$ includes any set for which $\Cvkvm$ is empty at a given time. The cardinality $|C|$ of configuration space is thus determined entirely by the support of $\pk$, along with $n$. Since $\vkvmclass$ does not convey ego state, just edge and neighbour configuration, we partition $\Cvkvm$ into sets of uninfected and infected nodes, such that $\Cvkvm = S_{\vkvm} \cup I_{\vkvm}$. Similar definitions allow us to introduce $S_{\vk}$ and $I_{\vk}$, $S_k$ and $I_k$, as well as $S$ and $I$. Although in general $|S_{\vkvm}| \neq |I_{\vkvm}|$, the structure of the uninfected and infected configuration spaces is identical, such that $|C| = |S| = |I|$, $|\Ck| = |S_{k}| = |I_{k}|$ and $|\Cvk| = |S_{\vk}| = |I_{\vk}|$.

The evolution of a dynamical process over a network amounts to a flow of nodes through the sets $\Svkvm$ and $\Ivkvm$ over time. Since the number of nodes $N$ in the network is conserved, it is their distribution over the sets $\Svkvm$ and $\Ivkvm$ that evolves in time. These distributions provide the \textit{state} of the network at time $t$.  Since our formalism is independent of network size, we deal with the densities of nodes rather than the absolute sizes of these sets. To this end we introduce
\begin{equation}\label{eqn:ck}
  \| \Ck \| \equiv \sum_{\Cvkvm \in \Ck} |\Cvkvm|
\end{equation}
as shorthand for the number of \textit{nodes} with underlying degree $k$. This is in contrast to $|\Cvk|$ and $|\Ck|$ which give the number of \textit{configurations} with degrees $\vk$ and $k$, respectively.  To convert from absolute node count to densities of nodes, we need to normalise $\Svkvm$ and $\Ivkvm$ by some non-zero quantity that is conserved over the course of a dynamical process. Since our temporal network models assume a static underlying network, a node's underlying degree $k$ is preserved, and as a result, so is $\| \Ck \|$, defined in \eref{eqn:ck}. The density of uninfected nodes in class $\vkvmclass$ in this case is given by
\begin{equation}
  \skm = \dfrac{|S_{\vkvm}|}{\| C_{k} \|},
\end{equation}
with $\ikm$ defined analogously. The node conservation principle leads to the condition $\sum_{C_k}(\skm + \ikm) = 1$, which is to say that the sum of all densities $\skm$ and $\ikm$ with underlying degree $k$, is one. We then have
\begin{equation}\label{eqn:rhokdeftemporal}
  \rho_{k} = 1 - \sum_{C_k} s_{\vkvm}
\end{equation}
and
\begin{equation}\label{eqn:rhodeftemporal}
    \rho = \sum_{k} p(k) \rho_{k},
\end{equation}
where the sum in the first expression is over all configurations $\vkvmclass$ that satisfy $\sum_j\kj = k$.  The term $\rho_k$ gives the probability that a randomly selected node with underlying degree $k$ will be infected, and $\rho$ the probability that any randomly selected node will be infected.  

As discussed in the main text, $\vs$ is the $|C|$-dimensional vector storing the densities $\skm$. In practice, we use lexicographic ordering of the tuples in $C$ to define a one-to-one mapping $\vkvmclass \mapsto i$, for some $i \in \{1, \hdots, |C| \}$ to define the $i$-th element $s_i$ of $\vs$. Finally, it is possible to show that for fixed $n$ and limiting $k$, the size of $C$ behaves like $\Theta (k^{2n})$. Now that we have defined the space of allowed configurations, we turn to its dynamics.


\textit{Master equation transition rates}. Ego transitions occur at rates $\Fkm$, and involve the flow of nodes from set $S_{\vkvm}$ to $I_{\vkvm}$. As such, no change to the ego's local neighbourhood $\vkvmclass$ takes place, and the transition represents a type of self-edge, or loop, in the lattice representation of configuration space, illustrated in \SNmaster{}. The rates $\Fkm$ are encoded in transmission functions such as those shown in \tref{tab:tab1}. Flux measurements of these transitions, such as those in \SNmeanfield{}, are expected to be exact, and are an important benchmark for verification of experiments. The rates $\Fkm$ are contained in the matrix $W_{ego}$.

Neighbour transitions are based on the probability $\beta_j dt$ that an uninfected neighbour of an uninfected node becomes infected over an interval $dt$. To calculate $\beta_j$ we use a straightforward ensemble average over $S$. To obtain the expected fraction of neighbours undergoing transitions, we observe the number of nodes undergoing ego transitions at time $t$, and count the number of neighbour transitions produced as a result. That is, when an uninfected node in class $\vkvmclass$ becomes infected, which occurs with probability $\Fkm dt$, it has $\kj -\mj$ uninfected neighbours that observe this transition, or $\kj - \mj$ nodes undergoing neighbour transitions.  The number of such edges across the entire network is given by $\sum_{S}\pvk (\kj - \mj) \Fkm \skm$, where the sum is over all uninfected classes. We compare this to the total number of uninfected-uninfected edges, $\sum_{S}\pvk (\kj - \mj) \skm$, giving the neighbour transition rate
\begin{equation}\label{eqn:betasj}
  \beta_j dt = \dfrac{\sum_{S}\pvk  (\kj - \mj) F_{\vk,\vm} s_{\vk,\vm}}
  {\sum_{S}\pvk (\kj - \mj) s_{\vk,\vm}} dt,
\end{equation}
which has previously been used in master equation solutions of binary-state dynamics on static networks. The rates $\beta_j$ are contained in the matrix $W_{neigh}$, weighted by the values $k_j$ and $m_j$ of the relevant classes $\vkvmclass$, as detailed in \SNmaster{}.

Edge transitions occur at rates $\mu_j$ and $\nu_j$, and give the probability of edges in state $j$ transitioning to state $j + 1$ or $j - 1$, respectively, over an interval $dt$. Their value depends upon the temporal network model in question. In this work, edge transition rates are determined by renewal processes following interevent time distributions $\psi (\tau)$, with complementary cumulative distributions $\Psi$. If the state of an edge is determined by the number of events $j$ having occurred in the preceding time window of duration $\eta$ due to a renewal process, edge transition rates are
\begin{equation}\label{methodeqn:mu}
    \mu_j dt = \dfrac{\Psi \ast \psi^{\ast j}}{\Psi \ast \psi^{\ast (j-1)} \ast \Psi} dt
\end{equation}
and
\begin{equation}\label{methodeqn:nu}
    \nu_j dt = \dfrac{\Psi \ast \psi^{\ast (j - 1)}}{\Psi \ast \psi^{\ast (j-1)} \ast \Psi} dt,
\end{equation}
with
\begin{equation}\label{methodeqn:E}
    E_j = \Psi \ast \psi^{\ast (j-1)} \ast \Psi
\end{equation}
giving the probability that a randomly selected edge is in state $j$. It is this quantity that provides the normalising constant for the rates $\mu_j$ and $\nu_j$. Here, $\psi^{\ast j}$ is the $j$-th convolution power of $\psi$. A complete derivation is given in \SNmaster{}. The Gaver-Stehfest algorithm is used to compute the inverse Laplace transforms, and an efficient numerical procedure reducing $\mu_j$ and $\nu_j$ to a matrix-vector product is developed in \SNlaplace{}. These expressions hold for $j > 0$, with \erefs{eqn:xiE}{eqn:ximu} in the main text giving the special case of $j = 0$ for $E_j$ and $\mu_j$, respectively.  Regardless of the form of $\psi$, the mean edge state $\eta / \langle \tau \rangle$ is always conserved on a network-wide level. Applying \erefs{methodeqn:mu}{methodeqn:nu} at the level of class transitions amounts to a mean field approximation, since flux measurements of Monte Carlo simulation show edge transition rates to deviate slightly from $\mu_j$ and $\nu_j$ at the class level $\vkvmclass$, even if exact for the network as a whole, as shown in \SNmeanfield{}. 


\textit{Simulation}. We simulate networks $\mathcal{G} = (\mathcal{V}, \mathcal{E})$ composed of a node set $\mathcal{V}$ of size $N$, and an underlying edge set $\mathcal{E}$. The edge set is produced by a desired degree distribution, wired according to the configuration model. Overlying temporal network activity is initialised to the steady state, such that at time $t = 0$, the time to the first event follows exactly the residual distribution $\Psi$, in the limit of large networks. Specifically, we set the time to $t = -\eta$, and draw $|\mathcal{E}|$ residual times from $\Psi$, or one for each edge. Subsequent interevent times are drawn from $\psi$. Advancing in time from $-\eta$ ensures that a stationary distribution of edge states $E_j$ is achieved \textit{exactly} at $t = 0$, when we begin to allow node dynamics to evolve. Due to the large values of interevent time standard deviation studied in this work, out-of-the-box sampling routines were either inefficient or broke down for large $\sigma_\tau$. As such, we develop a simple, yet efficient routine in \SNmontecarlo{} based on approximate inverse transform sampling of $\psi$ and $\Psi$, using a bisection method. This is performed on a numerical grid of $\Psi$ values, with relevant details of the probability distributions outlined in detail in \SNprobability{}. A third-order spline interpolation on a logarithmic scale provides intermediate values of the grid, such that the resultant underlying distribution is close to exact.

Node dynamics are implemented via a Gillespie algorithm, which uses the fact that the waiting time to infection for an uninfected node in class $\vkvmclass$ follows an exponential distribution with mean $1 / \Fkm$. Initially all nodes are in the uninfected state, and the diffusion process is triggered by low-level background noise at rate $p$. To simulate the temporal network itself, a time-ordered sequence of edge events is implemented in parallel with the node update sequence. This amounts to two separate time-ordered sequences of events executed simultaneously. Algorithms are described in detail in \SNmontecarlo{} with pseudocode.

We use the normalised density of noise-induced infections, $\rho_f$, and normalised diffusion time, $t_f$, as measures of the diffusion process. We define these quantities as follows. The probability that a randomly selected node has been infected as a result of external noise is
\begin{equation}
  \tilde{\rho}(t) = p \int_{0}^{t} (1 - \rho (\tau)) d\tau,
\end{equation}
meaning $0 < \tilde{\rho} \leq \rho$. We define $\rho_f$ as the fraction of infections that are due to noise $\rho_f = \tilde{\rho} / \rho$, such that $0 < \rho_f \leq 1$. This value cannot equal zero since there must be at least one noise induced infection, namely, the first infection in the diffusion process. A value approaching $\rho_f = 1$ means almost all infection is due to external noise. This occurs in the annealed limit, when almost all edges are in state $j = 0$, and network interactions play a vanishingly small role in the diffusion process. As a consequence, the time evolution of the diffusion process is governed by 
\begin{equation}
  \dot{\rho} = p(1 - \rho) 
\end{equation}
whose solution $\rho = 1 - e^{-pt}$ can be inverted to give the time required to the achieve a given density $\rho$ of infections relying solely on noise, that is,
\begin{equation}\label{eqn:tnorm}
  t = \dfrac{-\ln (1 - \rho)}{p}.
\end{equation}
If $t_c$ is the time required in the general case to reach a cutoff density of infections $\rho_c$, normalising $t_c$ by \eref{eqn:tnorm} evaluated at $\rho_c$ defines $t_f$, such that $0 < t_f \leq 1$. A value of $t_f = 1$ means the system is driven entirely by noise, and a value approaching $0$ a rapid diffusion process. An important feature of this work is that $t_f$ and $\rho_f$ seem to be interchangeable, as per the inset of \fref{fig:fig2}(a), and any result shown in terms of $\rho_f$ produces an identical picture in $t_f$.


\textit{Data description}. In this work we use two empirical temporal networks used by~\cite{saramaki2015exploring} and references therein, which we describe below. To simulate diffusion processes on these networks we use periodic boundary conditions, starting at a randomly selected point in time.

The first dataset is a temporal network of email exchange~\cite{saramaki2015exploring,eckmann2004entropy}, extracted from the log files of a university email server. The sender, recipient and the timestamp are used to form the network. The dataset consists of $N = 3188$ nodes, and $|\mathcal{E}| = 31857$ underlying edges, such that the average degree is $19.99$. A total of $308 730$ events were recorded, with a resolution of one second over a period of $81.3$ days. An average of $9.691$ events occur per edge. We determine the interevent time distribution by taking the the subset of edges observing more than one event, of which there are $21199$. The mean interevent time is then calculated to be $\langle \tau \rangle = 3.125 \text{ days}$, with standard deviation $\sigma_\tau = 6.620 \text{ days}$. This yields a coefficient of variation $\sigma_\tau / \langle \tau \rangle = 2.118$.

The second dataset is a temporal network of forum interactions~\cite{saramaki2015exploring,karimi2014structural}, an online community where users discuss movies. Similar to the email dataset, the sender, recipient and the timestamp are extracted from the messages. The dataset consists of $N = 7083$ nodes, and $|\mathcal{E}| = 138144$ underlying edges, such that the average degree is $39.01$. A total of $1428493$ events were recorded, with a resolution of one second over a period of $3133$ days. An average of $10.34$ events occur per edge. We determine the interevent time distribution by taking the the subset of edges observing more than one event, of which there are $70902$. The mean interevent time is then calculated to be $\langle \tau \rangle = 16.60 \text{ days}$, with standard deviation $\sigma_\tau = 76.53 \text{ days}$. This yields a coefficient of variation $\sigma_\tau / \langle \tau \rangle = 4.611$.

%% file: supplementary_preamble.tex
\makeatletter
\def\l@subsection#1#2{}
\def\l@subsubsection#1#2{}
\makeatother

\renewcommand{\tocname}{C\lowercase{ontents}}

\newcommand{\seref}[1]{Supplementary Eq.~(\ref{#1})}
\newcommand{\serefs}[2]{Supplementary Eqs.~(\ref{#1}) and (\ref{#2})}
\newcommand{\sfref}[1]{Supplementary Fig.~\ref{#1}}
\newcommand{\sfrefs}[2]{Supplementary Figs.~(\ref{#1}) and (\ref{#2})}
\newcommand{\stref}[1]{Supplementary Table~\ref{#1}}

\renewcommand\figurename{\text{Supplementary Figure}}
\renewcommand\thefigure{\text{\arabic{figure}}}

\renewcommand\tablename{\text{Supplementary Table}}

\newcounter{suppnote}
\refstepcounter{suppnote}

\newcommand{\SN}{\lowercase{\uppercase{S}upplementary \uppercase{N}ote}}

\newcommand{\MTmirror}{Fig. 1}
\newcommand{\MTheat}{Fig. 2}
\newcommand{\MTcollapse}{Fig. 3}

%% file: supplementary_v3.tex




\setcounter{page}{1}
\setcounter{figure}{0}
\setcounter{table}{0}
\setcounter{equation}{0}

\section*{\lowercase{\uppercase{C}ontents}}

\linespread{1.2}
\noindent
Supplementary Note 1. Master equation solution\hfill 1\\
Supplementary Note 2. Random walk equivalence\hfill 7\\
Supplementary Note 3. Edge-state distribution\hfill 7\\
Supplementary Note 4. Monte Carlo simulation\hfill 8\\
Supplementary Note 5. Laplace transform inversion\hfill 11\\
Supplementary Note 6. Diffusion speed and noise\hfill 12\\
Supplementary Note 7. Mean field approximation\hfill 12\\
Supplementary Note 8. Skewness and entropy\hfill 14\\
Supplementary Note 9. Probability distributions\hfill 15\\
\linespread{1.1}

\section*{\SN~\arabic{suppnote}.~M\lowercase{aster equation solution}}
\refstepcounter{suppnote}

In this Supplementary Note we detail the master equation solution used to solve for binary-state dynamics on our temporal network model. The general approach will be to assign one of a finite number of \textit{types} to each edge, and allow this quantity to evolve over time. To formulate a master equation solution, one defines a state space of allowed node configurations, which we term \textit{configuration space}. The second step is to define the allowed transitions between node configurations. The time evolution of a probability density over this state space amounts to a set of first-order differential equations, or rate equations, that among other things, provides the total density of infected nodes at a given time.

\subsection*{Configuration space}

As discussed in the main text, a network can be partitioned by the configurations $\vkvmclass$, where each node is assigned exactly one configuration, at any point in time. As a reminder, $\vk$ and $\vm$ are $n$-dimensional vectors storing $k_j$ and $m_j$, the number of neighbours along edges of type $j$, and the number of \textit{infected} neighbours along edges of type $j$, respectively. As a consequence, we have $0 \leq m_j \leq k_j$, with $1 \leq j \leq n$. Now consider a network of size $N$, following a degree distribution $p_k$, where $k = \sum_j k_j$ is the total degree, in a system allowing a maximum of $n$ edge states. We introduce $\Cvkvm$, the set of all nodes in the network with local configuration $\vkvmclass$.  Whereas $\Cvkvm$ is a set of \textit{nodes}, we introduce $\Ck$ to define the set of all \textit{sets} $\Cvkvm$ with total degree $k$, that is $\Ck = \lbrace \Cvkvm \mid \sum_j\kj = k \rbrace$.  Finally, $C$ is the set of all \textit{possible} sets $\Cvkvm$.  Provided a distribution of total degrees $\pk$, and edge dimension $n$, we define
\begin{equation}
  C = \{\vkvmclass \mid k \in \text{supp} (p_k) \enspace \text{and} \enspace \vzero \leq \vm \leq \vk\},
\end{equation}
which partitions the network at any given time.  This includes sets for which $\Cvkvm = \emptyset$ at a given time. The cardinality of this universal set, $|C|$, is determined by the support of $\pk$, in addition to $n$. Since $\vkvmclass$ does not convey ego state, just edge and neighbour configuration, we partition $\Cvkvm$ into uninfected and infected nodes, such that $\Cvkvm = S_{\vkvm} \cup I_{\vkvm}$. Similar definitions allow us to introduce $S_{\vk}$ and $I_{\vk}$, $S_{k}$ and $I_{k}$, as well as $S$ and $I$. Although in general $|S_{\vkvm}| \neq |I_{\vkvm}|$, the structure of the uninfected and infected configuration spaces is identical, such that $|C| = |S| = |I|$, $|\Ck| = |S_{k}| = |I_{k}|$ as well as $|\Cvk| = |S_{\vk}| = |I_{\vk}|$.

\begin{figure*}[t]
  \hspace*{-10mm}\includegraphics[scale=1]{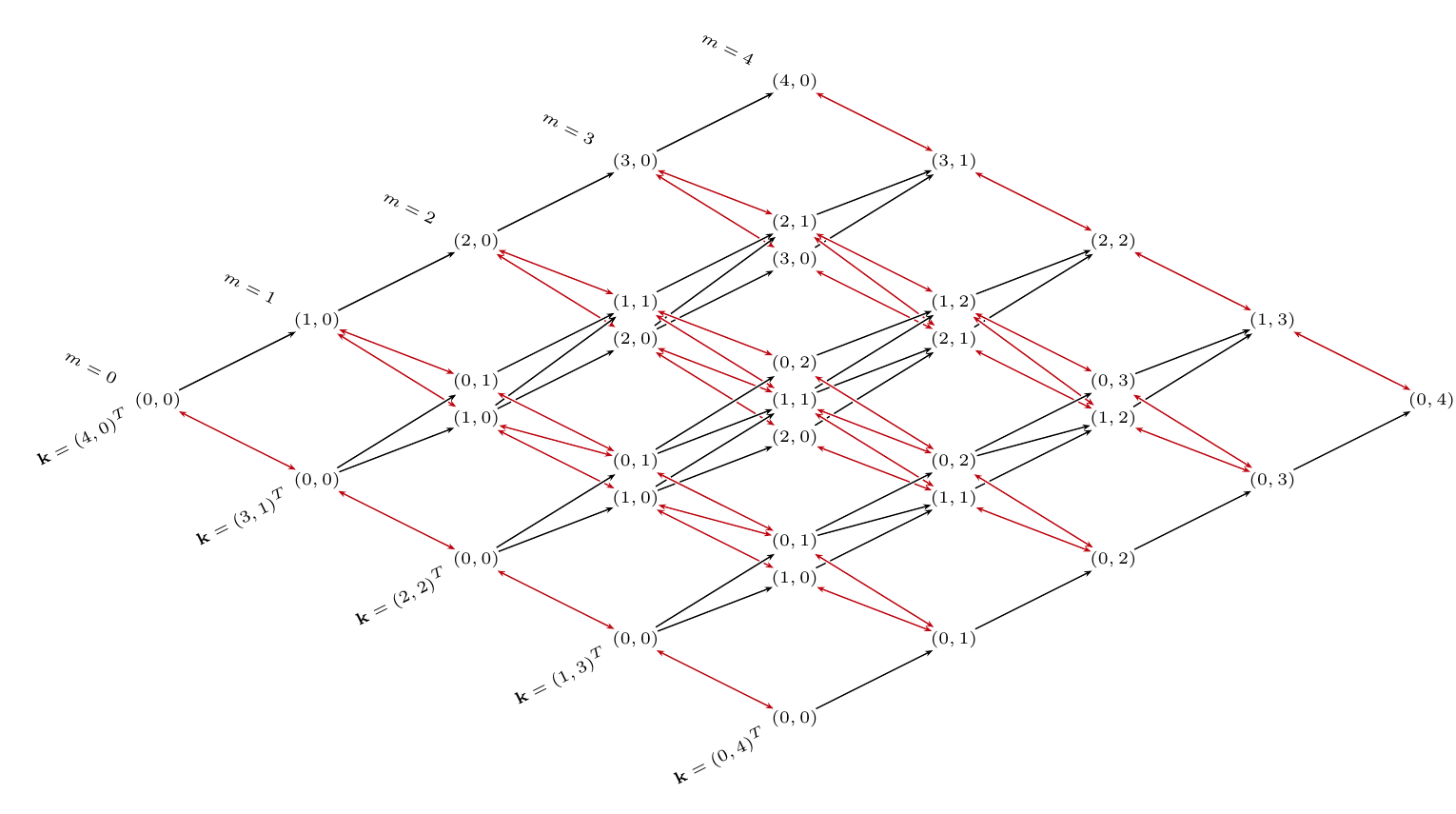}
  \caption[]{\textbf{Lattice diagram of temporal network configuration space.} Temporal network configuration space for a node of total degree $k = 4$, with $n = 2$ allowed edge states.  Right-diagonal transitions indicate neighbour infection, and left-diagonals the increments and decrements between edge-states $j = 0$ and $1$ that may occur in models of temporal networks.  Nodes in this lattice are labelled by their infected degree vector $\vm^{T}$, with corresponding $\vk$ shown.  Right-diagonal transitions preserve the degree vector $\vk$, indicated at the bottom of these diagonals.  Left-diagonal transitions preserve total infected neighbour count $m$, shown at the top of these diagonals. \label{fig:a1fig1}}
\end{figure*}

The evolution of a dynamical process over a network amounts to a flow of nodes through the sets $\Svkvm$ and $\Ivkvm$ over time. Since the number of nodes $N$ in the network is conserved, it is just their distribution over the sets $\Svkvm$ and $\Ivkvm$ that evolves in time.  These distribution provide the \textit{state} of the network at time $t$.  Since our formalism is independent of network size, we deal with the densities of nodes rather than the absolute sizes of these sets. As such, we define
\begin{equation}
  \| \Ck \| \equiv \sum_{\Cvkvm \in \Ck} |\Cvkvm|,
\end{equation}
in order to give the number of \textit{nodes} with degree vectors $\vk$ and $\vm$, and total degree $k$. This is in contrast to $|\Cvk|$ and $|\Ck|$ which give the number of \textit{configurations} with degrees $\vk$ and $k$.  To convert from absolute node count to densities of nodes, we need to normalise $\Svkvm$ and $\Ivkvm$ by some non-zero quantity that is conserved over the course of a dynamical process. For the temporal network models in question, the desired quantity is $\| \Ck \|$, defined above. The density of uninfected nodes in class $\vkvmclass$ in this case is given by
\begin{equation}
  \skm = \dfrac{|S_{\vkvm}|}{\| C_{k} \|},
\end{equation}
with $\ikm$ defined analogously. In the case of temporal networks, the node conservation principle leads to the normalisation condition $\sum_{\vkvm \mid k}(\skm + \ikm) = 1$, for a given $k$ class. We then have
\begin{equation}\label{eqn:rhodeftemporal}
  \rho_{k} = 1 - \sum_{\vkvm \mid k} s_{\vkvm}
\end{equation}
and
\begin{equation}\label{eqn:rhodeftemporal}
    \rho = \sum_{k} p_{k} \rho_{k},
\end{equation}
where the sum in the first expression is over all configurations $\vkvmclass$ that satisfy $\sum_j\kj = k$.  The time-dependent term $\rho_{k}$ gives the probability that a randomly selected node with total degree $k$ will be infected, and $\rho$ the probability that any randomly selected node will be infected. 

Finally, we define $\vs$ as the $|C|$-dimensional vector storing the densities $\skm$. In practice, we use lexicographic ordering of the tuples in $C$ to define a one-to-one mapping $\vkvmclass \mapsto i$, for some $i \in \{1, \hdots, |C| \}$ to define the $i$-th element $s_i$ of $\vs$. Finally, it is possible to show that for fixed $n$ and limiting $k$, the size of $C$ behaves like $\Theta (k^{2n})$.  Now that we have defined the space of allowed configurations, we turn to its dynamics.

\subsection*{Configuration transitions}

As outlined in the preceding section, the state of the system at time $t$ is given by the $|C|$ dimensional vector $\vs (t)$. After providing an initial condition $\vs (0)$, the evolution of the system can be approximated with the matrix $W (\vs, t)$, such that
\begin{equation}
  \dfrac{d}{dt}\vs = W(\vs, t)\vs = (W_{ego} + W_{neigh} + W_{edge})\vs,
\end{equation}
where $W$ can be decomposed into separate $|C|$-dimensional square matrices corresponding to flows driven by \textit{ego}, \textit{neighbour}, and \textit{edge} transitions, respectively. We outline these transitions in the following paragraphs. In the following we assume that besides the transitions specified, entries in $W_{ego}$, $W_{neigh}$ and $W_{edge}$ are zero. Despite $W$ being sparse, and the numerical implementation ultimately being in the form of dictionaries, we prefer the matrix form for exposition.

\begin{figure*}[t]
  \includegraphics[scale=1]{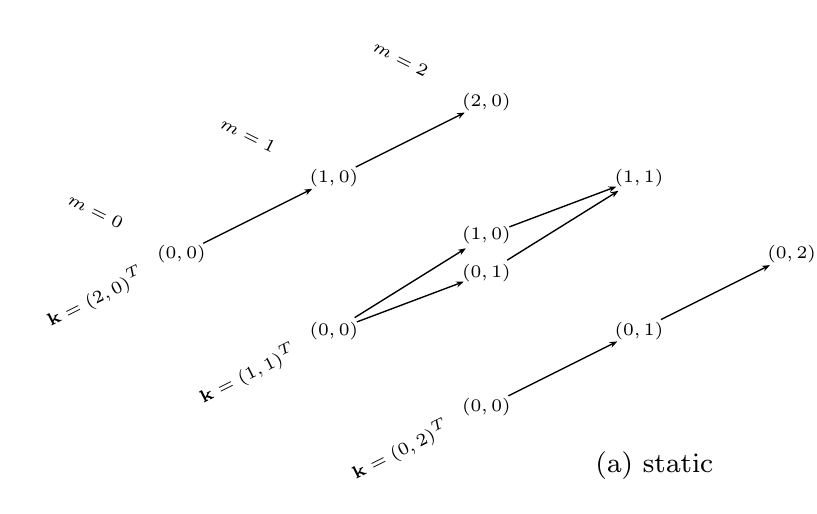}
  \hspace*{5mm}
  \includegraphics[scale=1]{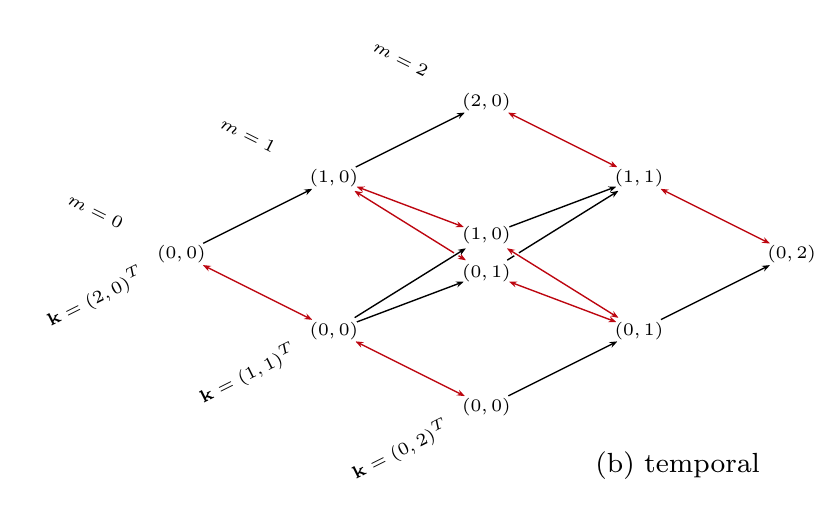}
  \caption{\textbf{Comparison of static and temporal configuration space.} Allowed transitions in static networks, (a), and temporal networks, (b), for a $k = 2$ degree node with an $n = 2$ level edge-state set. See caption in \sfref{fig:a1fig1} for interpretation. Note that the number of configurations $\vkvmclass$ is identical in each case, $|C_2| = 10$, it is just the number of allowed transitions that differ.\label{fig:a2fig1}}
\end{figure*}


\textit{Ego transitions}. In a non-recovery node dynamics, ego transitions of a node in class $\vkvmclass$ involve an uninfected node becoming infected, thereby exiting class $S$ and entering class $I$.  These transitions drive the actual node dynamics that overlie the temporal network substrate.  The transitions between these two classes are defined by infection and recovery rates $F_{\textbf{k}, \textbf{m}}$ and $R_{\textbf{k}, \textbf{m}}$ that depend on the dynamical model of interest.  Examples of $\Fkm$, the relative and absolute threshold rules, are given in the main text. We set $\Rkm = 0$ as we are interested in non-recovery dynamics here, where infected nodes cannot reenter the uninfected state, although this is straightforward to generalise. We assume that node transitions occur homogeneously in time regardless of the underlying dynamics. In other words, a node in the uninfected state becomes infected over an interval $[t, t + dt]$ with probability $F_{\textbf{k}, \textbf{m}}dt$. If $W(A, B)$ is the rate of transition from class $A$ to $B$, ego transitions in a non-recovery system are given by
\begin{equation}
  W(S_{\textbf{k},\textbf{m}}, I_{\textbf{k}, \textbf{m}}) = F_{\vk,\vm}.
\end{equation}
Since a node's egocentric network $\vkvmclass$ doesn't change during such a transition, the off-diagonal terms of $W_{ego}$ are zero, with the $i$-th diagonal term being $-F_i$, with ego transitions being a net loss to the $S$ set. In contrast, the following transitions correspond to off-diagonal matrices, since nodes undergoing these transitions remain in the $S$ class, and are compensated for elsewhere in $W$. These transitions appear as a type of self-loop in lattice diagrams of configuration space, \sfrefs{fig:a1fig1}{fig:a2fig1}.  Further, flux measurements of these transitions in Monte Carlo simulation ought to be exact, in the limit of large networks, and therefore act as a useful benchmark in transition rate studies. See \sfref{fig:figS5}(a) for an illustration.


\textit{Neighbour transitions}. Neighbor transitions refer to the change in nodes class due to the change in state of one of its neighbours, reflected in the value of the partial degree vector $\vm$.  We distinguish neighbour transitions by the type $j$ of the corresponding edge. The rates at which nodes leave the class $\vkvmclass$ due to neighbour infection are given by
\begin{equation}
  W(S_{\textbf{k},\textbf{m}}, S_{\textbf{k}, \textbf{m}+\textbf{e}_j}) = \beta_j(\kj - \mj).
\end{equation}
The coefficient $\beta_j$ gives the rates at which uninfected neighbours of uninfected nodes become infected. This quantity is derived below.  Influx to $\vkvmclass$ from the class $(\vk, \vm - \textbf{e}_j)$ due to the same mechanism is
\begin{equation}
  W(S_{\textbf{k},\textbf{m}-\textbf{e}_j}, S_{\textbf{k}, \textbf{m}}) = \beta_j(\kj - \mj + 1).
\end{equation}
To calculate $\beta_j$ we use a straightforward ensemble average, or mean-field approximation, over the set of all uninfected nodes. To obtain the expected fraction of neighbours undergoing transitions, we observe the number of egos undergoing transitions at time $t$, and count the number of neighbour transitions thus produced. That is, when an uninfected node in class $\vkvmclass$ becomes infected, which occurs with probability $\Fkm dt$, it produces $\kj -\mj$ uninfected nodes that observe neighbour transitions.  The number of such edges across the entire network is given by $\sum_{S}\pvk (\kj - \mj) \Fkm \skm$, where the sum is over all uninfected classes. We compare this to the total number of uninfected-uninfected edges, $\sum_{S}\pvk (\kj - \mj) \skm$, giving the neighbour transition rate
\begin{equation}\label{eqn:betasj}
  \beta_j dt = \dfrac{\sum_S \pvk  (\kj - \mj) F_{\vk,\vm} s_{\vk,\vm}}
  {\sum_S \pvk (\kj - \mj) s_{\vk,\vm}}dt,
\end{equation}
which has previously been used in master equation solutions of binary-state dynamics on static networks, see main text for references. At this point in the derivation, we could stop and write $W = W_{ego} + W_{neigh}$ in order to recover the static network transition matrix. These transitions appear as the right diagonals in lattice diagrams of configuration space, \sfrefs{fig:a1fig1}{fig:a2fig1}, where they preserve the degree vector $\vk$.  Further, flux measurements of these transitions in Monte Carlo simulation appear to show that \seref{eqn:betasj} is exact, in the limit of large temporal networks. See \sfref{fig:figS5}(b) and (c) for an illustration. 


\textit{Positive edge transitions}. The temporal nature of the underlying network is implemented using changes to edge states in the network.  A positive edge transition refers to the change in a node's configuration due to an increment in the state of one of its edges over an interval $dt$.  Regardless of the interpretation of edge state and the mechanism driving the transition, the probability of this occurring on a randomly selected edge of type $j$ is $\mu_j dt$. In fact, we delay until the next section our discussions of models of temporal networks - for now it suffices to assume that they can be represented by networks with dynamic edge state.  For a configuration $\vkvmclass$, a positive edge transition on a $j$-type edge means losing an edge of that type, and gaining an edge of type $j + 1$. For brevity, we introduce the term $\Delta_j^\pm = -\textbf{e}_j + \textbf{e}_{j \pm 1}$, corresponding to the change in the degree vector $\vk$ imposed by such a transition.  That is, an adjacent node loses a $j$-type edge, and gains a $j \pm 1$-type edge, all while preserving the underlying degree $k$.  The symmetry relations $\Delta_j^+ = -\Delta_{j + 1}^-$ and $\Delta_j^- = -\Delta_{j - 1}^+$ clearly hold.

The configuration that a $\vkvmclass$ node enters when undergoing a positive transition on a $j$-type edge is $(\vk + \Delta_j^+,\vm)$, or $(\vk + \Delta_j^+,\vm + \Delta_j^+)$, depending on whether the neighbouring node was uninfected or infected, so that we have
\begin{equation}
  W(S_{\textbf{k},\textbf{m}}, S_{\textbf{k} + \Delta_j^+,\textbf{m}}) = \mu_j (\kj - \mj)
\end{equation}
and
\begin{equation}
  W(S_{\textbf{k},\textbf{m}}, S_{\textbf{k} + \Delta_j^+,\textbf{m} + \Delta_j^+}) = \mu_j \mj,
\end{equation}
respectively.  Similarly, nodes may enter the configuration $\vkvmclass$ through a positive transition on a $j - 1$ edge, via the classes $(\vk - \Delta_j^-, \vm)$ and $(\vk - \Delta_j^-, \vm - \Delta_j^-)$. We have
\begin{equation}
  W(S_{\textbf{k} - \Delta_j^-,\textbf{m}}, S_{\textbf{k},\textbf{m}}) = \mu_j (\kj - \mj + 1)
\end{equation}
and
\begin{equation}
  W(S_{\textbf{k} - \Delta_j^-,\textbf{m} - \Delta_j^-}, S_{\textbf{k},\textbf{m}}) = \mu_j (\mj + 1),
\end{equation}
if the neighbour is uninfected or infected, respectively.  Combining these terms gives the flow through the configuration $\vkvmclass$ due to positive transitions on $j$-type edges. Note that we typically impose boundary conditions, if not because sharp cutoffs arise naturally in many temporal network models, because we require the configuration space to remain finite.  If $n$ remains the number of allowed edge states, we impose the condition that one cannot observe a positive edge transition on an $n$-type edge.  As such, a node cannot lose an $n$-type edge through a positive edge transition, and we write $\mu_n = 0$. 


\textit{Negative edge transitions}. A negative edge transition refers to the change in a node's configuration when an adjacent event is forgotten over an interval $dt$, causing a decrease in the number of memorable events on that edge. As defined above, this occurs with probability $\nu_j dt$. When an event terminates on an edge of type $j$, it gains an edge of type $j - 1$, and loses and edge of type $j$, preserving the total degree $k$.  The relations between node classes due to negative edge transitions mirror their positive counterparts, and we include them here for completeness.  A node in configuration $(\vk,\vm)$ moves to class $(\vk + \Delta_j^-,\vm)$ and $(\vk + \Delta_j^-,\vm + \Delta_j^-)$, if an event on a $j$-type edge terminates while connected to an uninfected or infected neighbour, respectively.  This occurs at rates
\begin{equation}
  W(S_{\textbf{k},\textbf{m}}, S_{\textbf{k} + \Delta_j^-,\textbf{m}}) = \nu_j (\kj - \mj)
\end{equation}
and
\begin{equation}
  W(S_{\textbf{k},\textbf{m}}, S_{\textbf{k} + \Delta_j^-,\textbf{m} + \Delta_j^-}) = \nu_j \mj .
\end{equation}
Similarly, nodes may enter the configuration $\vkvmclass$ with a negative edge transition on a $j + 1$ edge, from the classes $(\vk - \Delta_j^+, \vm)$ and $(\vk - \Delta_j^+, \vm - \Delta_j^+)$ as follows. If the event terminates between a node and an uninfected neighbour, we have
\begin{equation}
  W(S_{\textbf{k} - \Delta_j^+,\textbf{m}}, S_{\textbf{k},\textbf{m}}) = \nu_j (\kj - \mj + 1)
\end{equation}
and
\begin{equation}
  W(S_{\textbf{k} - \Delta_j^+,\textbf{m} - \Delta_j^+}, S_{\textbf{k},\textbf{m}}) = \nu_j (\mj + 1),
\end{equation}
if the neighbour is infected.  Combining these terms gives the flow through the configuration $(\vk,\vm)$ due to negative edge transitions. Note the boundary condition, namely that a $j = 0$ edge is the case where no events have taken place in the last $\eta$ interval. As such, a node cannot lose a $0$-type edge through a negative edge transition, and we reflect this by writing $\nu_{0} = 0$.  These transitions appear as left diagonals in lattice diagrams of configuration space, \sfrefs{fig:a1fig1}{fig:a2fig1}, and preserve the total number of infected neighbours $m$. Further, flux measurements of these transitions in Monte Carlo simulation show that constant $\mu_j$ and $\nu_j$ can be excellent approximations of non-Markovian systems. See \sfref{fig:figS5}(d) to (f) for an illustration.

\subsection*{Calculating $\mu_j$ and $\nu_j$ for renewal processes}\label{sec:Emunuc}

\begin{figure}
  \centering
  \includegraphics[scale=1]{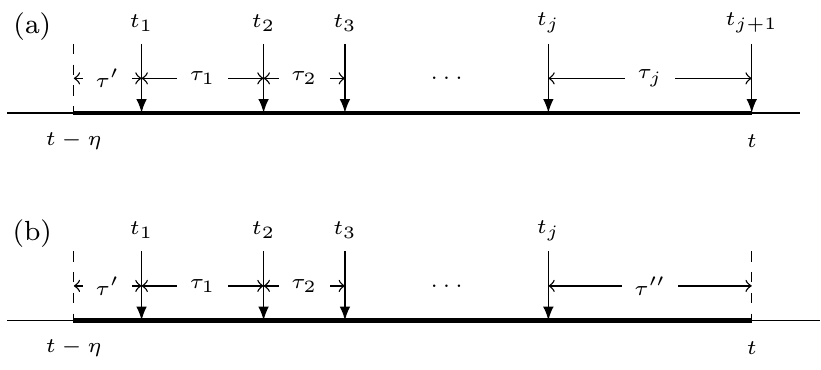}
  \caption{\textbf{Enumeration of edge state configurations.} Configuration of $j$ events occurring over an interval of length $\eta$, where $\eta$ amounts to memory of observed events.  In plot (a) we enumerate all configurations of edges in state $j$ at time $t$, with a $(j + 1)$-th event occurring over the interval $t < t_{j + 1} < t + dt$. In plot (b) we enumerate all the configurations of edges in state $j$. We use this to calculate the rate of positive edge transition $\mu_j$, and edge state distribution $E_j$, respectively. Interevent times $\tau_1, \hdots, \tau_j$ are drawn from $\psi$, whereas $\tau^{\prime}$ and $\tau^{\prime \prime}$ are drawn from $\Psi$, defined in the text. \label{fig:figS4}}
\end{figure}

In this section, we calculate the rates of positive and negative edge transitions $\mu_j$ and $\nu_j$ in the stochastic temporal network model discussed in the previous section. Here, $\mu_j dt$ gives the probability that at time $t$, for a renewal process having already produced $j$ events in the preceding time window of duration $\eta$, a $(j + 1)$-th event is observed between time $t$ and $t + dt$. Conversely, $\nu_j dt$ gives the probability of an event exiting the $\eta$ window. This is illustrated in \sfref{fig:figS4}. At the outset, describing such a process with constant rates $\mu_j$ and $\nu_j$ seems inappropriate, as it is memoryless only in so-called \textit{event space}.  In this representation, a renewal process is nothing other than a sequence of trials, $\{\tau_1, \tau_2, \hdots \}$, or the random sampling of a value $\tau$ from a distribution $\psi (\tau)$. In this sense, the process is memoryless.  However, in the resulting time-series for continuous $t$, the process is non-Markovian, as the time of the next event was decided at the time of the previous event, and the probability of an event occurring at a time between these points is zero. This is in contrast to a Poisson process, where the change in edge state due to the occurrence of an event is constant in time.  Thus, on a microscopic level, where we observe a stochastic process on a single edge, a rate description is nonsensical. 

We note, however, that in our master equation formalism, classes $\vkvmclass$ really represent ensembles of nodes, and although constant rates cannot be identified on a microscopic level, useful quantities do exist on a network-wide macroscopic level. That is, calculating the fraction of $j$-type edges that change state over an interval $dt$ turns out to be strongly heterogeneous for varying $j$, which is clearly not the case for a Poisson process. The heterogeneity of the distribution $\psi$ is reflected in the heterogeneity of $\mu$, $\nu$ and $E$.

The rate of positive edge transition $\mu_j dt$ is calculated by finding the probability of a $(j + 1)$-th event occurring over a given interval $dt$, on the condition that $j$ events have already been produced in the preceding time interval of duration $\eta$. This is illustrated in \sfref{fig:figS4}(a), where we set $t = \eta$ for convenience. In \sfref{fig:figS4}, the interevent times $\tau_1, \hdots, \tau_j$ are drawn from the distribution $\psi$, and the times $\tau^{\prime}$ and $\tau^{\prime \prime}$ from its complementary cumulative distribution $\Psi$, also known as the residual time distribution. It is defined as
\begin{equation}\label{eqn:psi_psi}
  \Psi(\tau) = \int_{\tau}^{\infty}\psi(t)dt,
\end{equation}
and gives the probability that the time between events is of duration at least $\tau$. We introduce the domain $T$ of times spanned by the configurations allowed in \sfref{fig:figS4}, or the times $t_1 < t_2 < ... < t_j$ in an interval of duration $\eta$, such that $t_j - t_1 < \eta$ and $t_{j + 1} - t_1 < \eta$. Note that consecutive event times $t_j$ and $t_{j + 1}$ cannot coincide, with interevent times drawn from distributions $\psi$ and $\Psi$, which defined over positive $\tau$. We write
\begin{align}
T =  & \left(0, \eta \right] \times \left(0, \eta - t_1 \right] \times \left(0, \eta - t_2 \right] \times \hdots \nonumber \\
   & \quad \hdots \times \left(0, \eta - t_{j - 2} \right] \times \left(0, \eta - t_{j - 1} \right] \subseteq  \textbf{R}^j_{+},
\end{align}
where $\textbf{R}^j_+$ is the $j$-dimensional space of positive real numbers. The probability of observing the configuration in \sfref{fig:figS4}(a) is $\Psi(\tau^{\prime})\psi(\tau_{1}) \hdots \psi(\tau_j)$, which is the same as $\Psi(t_1)\psi(t_2 - t_1)\hdots \psi(\eta - t_j)$ given that we've set the time $t$ to $\eta$ for simplicity. Similarly, the configuration in \sfref{fig:figS4}(b) is observed with probability $\Psi(\tau^{\prime}) \psi(\tau_{1}) \hdots \psi(\tau_{j - 1})\Psi(\tau^{\prime \prime})$, which is the same as $\Psi(t_{1})\psi(t_{2} - t_{1}) \hdots \psi(t_j - t_{j - 1})\Psi(\eta - t_j)$. The weighted sum of all such configurations yields the probability of observing a $j$ type edge undergoing a transition to state $j + 1$ over an interval $dt$, and the probability of randomly selecting an edge in state $j$, as shown in \sfref{fig:figS4}(a) and (b), respectively. The only difference is the final term, which is drawn either form $\psi$ or $\Psi$. With respect to the domain $T$, these sums can be written
\begin{align}\label{eqn:munum}
  \int_T {d}t_1 & \hdots {d}t_j \Psi(t_1)\psi(t_2 - t_1)\psi (t_3 - t_2) \times \hdots \nonumber \\
  & \hdots \times \psi (t_j - t_{j - 1})\psi(\eta - t_j) = \Psi \ast \psi^{\ast j},
\end{align}
and
\begin{align}\label{eqn:muden}
  & \int_{T} dt_1 \hdots dt_j \Psi(t_1) \psi(t_2 - t_1)\psi(t_3 - t_2)\times \hdots \nonumber \\
    & \hdots \times \psi(t_j - t_{j - 1}) \Psi(\eta - t_j) = \Psi \ast \psi^{\ast (j - 1)} \ast \Psi,
\end{align}
respectively. Here, $\psi^{\ast j}$ is the $j$-th convolution power of $\psi$, and is discussed at length in following sections. To obtain the rate $\mu_j$ at which edges in state $j$ transition to state $j + 1$, \seref{eqn:munum} must be normalised by \seref{eqn:muden}, the probability $E_j$ that a randomly selected edge is in state $j$. Schematically, this corresponds to normalising the transition in \sfref{fig:figS4}(a) by those in \sfref{fig:figS4}(b). Note that $\nu_j dt$, the probability that an edge in state $j$ forgets an event over an interval $dt$ is the same as $\mu_{j - 1}$ under time reversal, up to the normalising constant. As such, the rates $\mu_j$ and $\nu_j$, along with the distribution $E_j$, can be written compactly as
\begin{equation}\label{eqn:mu}
  \mu_j dt = \dfrac{\Psi \ast \psi^{\ast j}}{\Psi \ast \psi^{\ast (j-1)} \ast \Psi} dt
\end{equation}
and
\begin{equation}\label{eqn:nu}
  \nu_j dt = \dfrac{\Psi \ast \psi^{\ast (j - 1)}}{\Psi \ast \psi^{\ast (j-1)} \ast \Psi} dt,
\end{equation}
with
\begin{equation}\label{eqn:E}
  E_j =  \Psi \ast \psi^{\ast (j-1)} \ast \Psi.
\end{equation}
These quantities  can be calculated either numerically or analytically, depending on the tractability of the chosen distribution $\psi$.  In general, if $\psi(\tau)$ is locally integrable, then the Laplace transform of $\psi$ and $\Psi$ exists and allows us to calculate the convolution as a product in the frequency domain, which will be useful especially if $j$ is large. Since we don't impose any cutoffs on $\psi$ and $\Psi$ in the text, $j$ indeed can grow arbitrarily large, under bursty dynamics.  If we denote the Laplace transform of $\psi(\tau)$ by 
\begin{equation}\label{eqn:psi_hat}
  \mathcal{L}\left\lbrace \psi(\tau) \right\rbrace = \hat{\psi}(s) = \int_{0}^{\infty}\psi(\tau)e^{-s\tau}d\tau
\end{equation}
then the transform of \seref{eqn:psi_psi} can be written as
\begin{equation}\label{eqn:Psi_hat}
  \hat{\Psi}(s) = \dfrac{1 - \hat{\psi}(s)}{s}.
\end{equation}
Finally, we can argue that by induction from $j = 0$, and using the fact that the distribution $E$ is constant at stationarity of the renewal process, that $\mu_j E_j = \nu_{j + 1} E_{j + 1}$, meaning that along with $E_j$, the rates $\mu_j$ and $\nu_j$ are stationary.  The observed experimental rates $\mu_j$, $\nu_j$ and $E_j$ match exactly the predicted values, for increasingly large networks. 

\subsection*{Illustration using the exponential distribution}

In this section we explicitly calculate the edge transition rates $\muj$ and $\nuj$ for an exponential interevent time distribution $\psi$. This is an exercise to illustrate the Laplace inversion procedure, in general we calculate these quantities numerically, as described in \SNlaplace. The exponential distribution is an important benchmark in this work, and we consider two alternative generalisations, namely the gamma and Weibull distributions. Both reduce to the exponential distribution when $\sigma_\tau = \langle \tau \rangle = 1$, and recover the rates given here. Consider such a distribution with average $\langle \tau \rangle$ defined by
\begin{equation}\label{eqn:psi_exponential}
	\psi(\tau) = \dfrac{1}{\langle \tau \rangle} e^{-\tau / \langle \tau \rangle}
\end{equation}
with
\begin{equation}
	\Psi(\tau) = e^{-\tau / \langle \tau \rangle},
\end{equation}
having transforms
\begin{equation}
  \hat{\psi}(s) = \dfrac{1}{\langle \tau \rangle s + 1}
\end{equation}
and 
\begin{equation}
  \hat{\Psi}(s) = \dfrac{1 - \hat{\psi}}{s} = \dfrac{\langle \tau \rangle}{\langle \tau \rangle s + 1},
\end{equation}
respectively. Substituting these transforms into \seref{eqn:E}, and applying the convolution theorem allows us to calculate the expected size of the set of edges in state $j$, which is also the normalising constant in the rates $\mu_j$ and $\nu_j$, as the expression for $\mathcal{L}\left\lbrace E_j \right\rbrace$ simplifies to to
\begin{equation}
  \hat{\Psi} \cdot \hat{\psi}^{j-1} \cdot \hat{\Psi} = \dfrac{\tauavg^2}{\left(\tauavg s + 1\right)^{j + 1}} .
\end{equation}
We use the fact that $\mathcal{L}^{-1}\{ \tfrac{1}{s^{j + 1}} \} = \eta^j / j!$, for integer $j$, a known Laplace transform relating to the gamma function. We use also the translation property $\mathcal{L}^{-1} \{\hat{\psi}(s + a)\} = e^{-a}\psi$, which directly results from the definition \seref{eqn:psi_hat}. The inverse Laplace transform of the above expression can then be written explicitly as
\begin{equation}\label{eqn:mu_den}
 \Psi \ast \psi^{\ast (j-1)} \ast \Psi =  \tauavg \dfrac{e^{-\eta / \langle \tau \rangle}}{j!}\left( \dfrac{\eta}{\langle \tau \rangle} \right)^j, 
\end{equation}
meaning $E$ is simply the Poisson distribution with mean $\eta / \langle \tau \rangle$. This is expected, due to our construction of the memory window, and the fact that an exponential interevent time distribution recovers a Poisson process. Similarly, the Laplace transform of the numerator in \serefs{eqn:mu}{eqn:nu} can be used to calculate
\begin{equation}\label{eqn:mu_num}
   \Psi \ast \psi^{\ast j} = \dfrac{e^{-\eta / \langle \tau \rangle}}{j!}\left( \dfrac{\eta}{\langle \tau \rangle} \right)^j
\end{equation}
and
\begin{equation}\label{eqn:mu_num}
   \Psi \ast \psi^{\ast (j - 1)} = \dfrac{e^{-\eta / \langle \tau \rangle}}{(j - 1)!}\left( \dfrac{\eta}{\langle \tau \rangle} \right)^{j - 1},
\end{equation}
yielding
\begin{equation}
  \mu_j dt = \dfrac{1}{\langle \tau \rangle}dt
\end{equation}
and
\begin{equation}
  \nu_j dt = \dfrac{j}{\eta}dt,
\end{equation}
after normalising by $\Psi \ast \psi^{\ast (j-1)} \ast \Psi$.  In this special case of exponentially distributed $\tau$, we are able to derive these rates using much simpler arguments, namely with the definition of the Poisson process, and the Poisson distribution. Crucially, $\mu_j$ has no $j$ dependence here, which clearly expresses the memoryless property of the Poisson process.  These rates may be verified by simulating an ensemble of independent, stationary renewal processes, and observing the flux in the system over an interval $\Delta t$.  The flow through the set $E_j$ over that interval, scaled by the size of that set and the size of the measurement window, give the rates $\mu_j$ and $\nu_j$.  Alternatively, by simulating a single renewal process for a sufficiently long time, and measuring its change in behaviour over each interval $\Delta t$, one obtains rates $\mu_j$ and $\nu_j$ that are identical to those calculated in the ensemble.

\subsubsection*{Convolution powers, an aside}

In order to calculate the distribution of edge states $E_j$, as well as the mean field edge transition rates $\mu_j$ and $\nu_j$, we need an efficient method for computing convolution powers. In general a convolution is defined for two real valued functions $f$ and $g$ over the domain $f, g : (-\infty, \infty) \rightarrow \mathbb{R}$. However, in the case where $f$ and $g$ take non-negative values, as is the case in our study, the convolution reduces to  
\begin{equation}
  (f \ast g)(t) = \int_{-\infty}^{\infty} f(\tau) g(t - \tau) d\tau = \int_{0}^{t} f(\tau) g(t - \tau) d\tau.
\end{equation}
We use this to express edge-state properties, \serefs{eqn:munum}{eqn:muden}, as convolutions over the non-negative reals. It is worthwhile noting the convention that if $\psi^{\ast j}$ is the $j$-th convolution power, or 
\begin{equation}
    \psi^{\ast j} = \underbrace{\psi \ast \psi \ast \hdots \ast \psi}_{\text{$j$ terms}},
\end{equation}
then the zeroth order convolution, $\psi^{\ast 0} = \delta_{0}$, is simply the Dirac delta function, the identity of convolution. This is used in the $j = 1$ case of Eqs. \ref{eqn:mu}, \ref{eqn:nu} and \ref{eqn:E}.

\section*{\SN~\arabic{suppnote}.~R\lowercase{andom walk equivalence}}
\refstepcounter{suppnote}

In this Supplementary Note, we discuss the expected steady-state network topology that emerges due to our stochastic temporal model. In particular, we describe the existence of a pure Markovian system with identical macroscopic dynamics to our non-Markovian renewal process model. We consider the dependence of the edge transition rates $\mu_j$ and $\nu_j$ on our choice of interevent time distribution $\psi$, and in particular, its parameterisation in terms of standard deviation $\sigma_\tau$. 


\begin{figure}
  \includegraphics[scale=1]{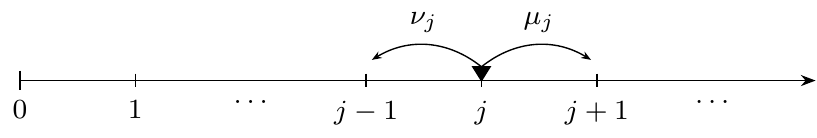}
  \caption{\textbf{Random walk interpretation of edge state.} Positive and negative edge transition rates, $\mu_j$ and $\nu_j$, act as a signature of the non-Markovianity in the renewal process model. On a macroscopic level,  this model is indistinguishable from a random walk as illustrated above, where the transition rates are provided by $\mu_j$ and $\nu_j$ by construction. In contrast, this equivalence is broken when taking into account node dynamics on the level of classes $\vkvmclass$, as we shown in \sfref{fig:figS5}. See also \sfref{fig:figS7} for illustrative values of the rates $\mu_j$ and $\nu_j$ for varying $\sigma_\tau$. \label{fig:figS9}}
\end{figure}

\begin{figure*}
  \includegraphics[scale=1]{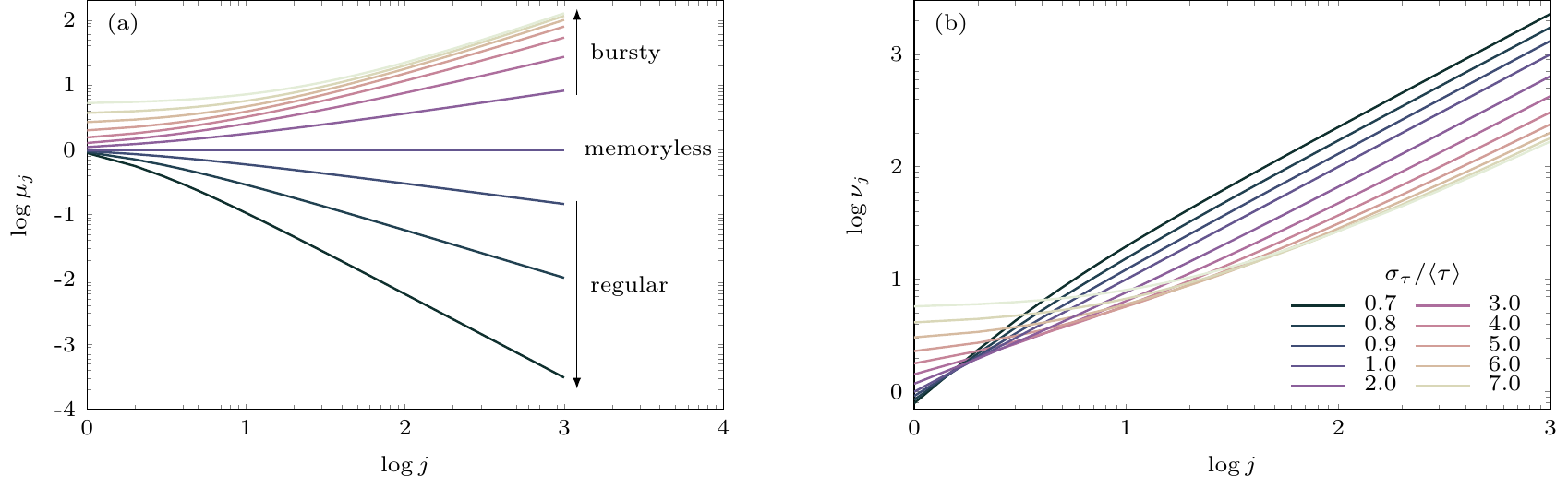}
  \caption{\textbf{Illustration of edge transition rates.} Positive and negative edge transition rates, $\mu_j$ and $\nu_j$, respectively, for a gamma distributed interevent time with mean $\langle \tau \rangle = 1$. On a macroscopic level, our renewal process model is indistinguishable from a random walk model of edge state, where the state $j$ increments and decrements at rates $\mu_j$ and $\nu_j$. Legend in (b) applies also to (a). Observer memory and mean interevent time are given by $\eta = \langle \tau \rangle = 1$, according to our stochastic temporal network model. The exponential distribution is recovered in the case of $\sigma_\tau = \langle \tau \rangle = 1$, producing a memoryless system, as indicated. \label{fig:figS7}}
\end{figure*}

We assume an arbitrarily large network consisting of independent, stationary renewal processes, such that the time to the next event at any time $t$ follows the residual distribution $\Psi (\tau)$. We illustrate in \sfref{fig:figS7} the edge transitions rates $\mu_j$ and $\nu_j$ that emerge from a gamma interevent time distribution $\psi (\tau)$, for increasing values of standard deviation $\sigma_\tau$. Since $\mu_j$ and $\nu_j$ are heterogeneous, they can be interpreted as providing a signature of the non-Markovianity inherent to the renewal process microscopically. For instance, when $\sigma_\tau = \langle \tau \rangle = 1$, the gamma distribution reduces to an exponential distribution, corresponding to a Poisson process. That this process is memoryless is reflected in the edge transition rate $\mu_j = 1 / \langle \tau \rangle$, being homogeneous in $j$. The greater the departure from the Poisson process, the greater the heterogeneity in $j$, illustrated in \sfref{fig:figS7}(a).

A central result in this work is to note that for an uncorrelated ensemble of edges, the renewal process model is indistinguishable from a continuous-time Markov chain, namely, a one-dimensional biased random walk, as illustrated in \sfref{fig:figS9}. In other words, the probability of a randomly selected edge undergoing a transition over an interval $dt$ in the renewal process model is trivially identical to a Markov chain where by construction, transitions occur at rates $\mu_j$ and $\nu_j$. Indeed, since all temporal network information is stored in $\mu_j$, $\nu_j$ and $E_j$ in the master equation, any class of system producing a given set of $\mu_j$, $\nu_j$ and $E_j$ values has the same predicted dynamics. Since in the special case of a Markov chain transition rates are exact at all scales (both locally on the scale of a single edge, and globally on the scale of an uncorrelated ensemble), the master equation solution is exact here. Since \serefs{eqn:mu}{eqn:nu} represent a mean field approximation on the scale of classes $\vkvmclass$ in the renewal process model, deviations in the master equation solution emerge here. These errors provide a measure of the extent to which non-Markovian dynamics can be captured by the heterogenity of a simple Markov chain. 

\section*{\SN~\arabic{suppnote}.~E\lowercase{dge-state distribution}}
\refstepcounter{suppnote}

In this Supplementary Note we mention some basic properties of the edge state distribution. First, for all choices of $\psi (\tau)$ and $\eta$ in this work, a useful conserved quantity is the expected edge state, or the expected number of events per edge across the entire network. If $\langle \tau \rangle$ is the mean of this distribution, and $\eta$ is observer memory, the expected edge state is $\langle E \rangle = \eta / \langle \tau \rangle$, and is useful for monitoring the accuracy of the implementation. Further, note that if $\mathcal{E}$ is the set of underlying edges in the network, the superposition of $|\mathcal{E}|$ renewal processes converges to a exponential distribution with mean $\langle \tau \rangle / |\mathcal{E}|$, for large $\mathcal{E}$.  As such, we expect $|\mathcal{E}|dt / \langle \tau \rangle$ events per time window $dt$ in simulation.

\subsection*{Temporal percolation transition}

As discussed in the main text, the probability that a randomly selected edge is in state zero is $\xi_E$. One can use $\xi_E$ to determine whether active edges in the network, the set of all edges in state one or higher, are expected to form a giant component at any given time. If a giant component exists, we say that percolation has taken place. If percolation does not occur, nodes form finite active clusters whose size goes to zero in the limit of large networks.  This is true even if the underlying network itself consists of a giant component. The percolation transition  helps to explain the sharp transition in dynamics in the upper right corners of \MTheat{} in the main text, separating regimes of fast and slow information diffusion.  The percolation condition for a configuration model network with degree distribution $q(k)$ is given by
\begin{equation}
  \sum_{k = 0}^{\infty} k (k - 2) q(k) = 0 ,
\end{equation}
where 
\begin{equation}
    q(k) = \sum_{l \geq k}^{\infty} p(l) \binom{l}{k} (1 - \xi_E)^k \xi_E^{l - k},
\end{equation}
is the degree distribution obtained by randomly removing a fraction $\xi_E$ of edges from a network with degree distribution $p(k)$, which is effectively the network induced when removing edges in state zero in out temporal model.

\subsection*{Maximum edges state}

In our analytic solution, we introduce $n$ to denote the maximum edge state $j$. In experiment, no such restrictions are imposed when sampling the interevent time distributions $\psi$ and $\Psi$, in contrast to related work~\cite{jo2014analytically} where it is common to introduces upper lower bounds on $\tau$.  As a consequence, bursts in activity can lead to arbitrarily large edge states $j$. However, as we see in the structure of configuration space, if $n$ is the maximum edge state allowed in the system, the number of equations grows like $\Theta (k^{2n})$. Clearly, in the interest of the numerical implementation of the master equation solution, $n$ cannot be arbitrarily large.

It is noteworthy that despite edge state being unrestricted in simulation, the resultant diffusion dynamics can be accurately solved using relatively small values of $n$. Consider that large bursts are most common when the interevent time standard deviation $\sigma_\tau$ is large. In this regime, the fraction of edges in state $j = 0$ is significant. As a consequence, nodes observing bursts of activity on some edges frequently observe no activity on others. Indeed for large enough $\sigma_\tau$, it is rare for a node to observe more than on active neighbour, with that active neighbours generally being in a very large state $j$. For the relative threshold (RT) model, such a node is infected with high probability if the neighbouring node is active, since the threshold $\phi$ is guaranteed to be overcome here. Similarly, for the absolute threshold (AT) model, any infected neighbour in state $j = \lceil M_\phi \rceil$ or higher is likely to adopt. Finally, for the susceptible-infected (SI) model, consider that the duration of a spike in activity is on the order of $\eta$, i.e., the duration of observer memory. Since the infection rate increases proportionally to the size of the burst, a large burst leads to infection soon after its observation, and long before they have left the time window. Since the local neighbourhood of the newly infected node is likely sparse in this setting, a much smaller burst would lead to an identical diffusion outcome. Effects such as these mean that surprisingly low value of $n$ are sufficient to accurately model the diffusion process.

\section*{\SN~\arabic{suppnote}.~M\lowercase{onte} C\lowercase{arlo simulation}}
\refstepcounter{suppnote}

In this Supplementary Note we discuss the Monte Carlo simulation methods used in this work.  Since node dynamics do not feed back into edge dynamics, we assume a steady state renewal process is in place at $t = 0$. This involves initialising the system with one $\tau$ drawn from the tail distribution $\Psi$. We start the simulation at $t = -\eta$, so that at time $t = 0$, the system is at steady state.

\subsection*{Parallel event sequences}

Our model of node dynamics is Markovian, since an uninfected node $v$ becomes infected at a constant rate $F_v$. In the case of threshold models, the transmission rate is a step function, whose upper value is set conventionally to one. A straightforward Monte Carlo simulation in this case is to advance in time with fixed intervals $\Delta t = \tfrac{1}{N}$, where $N$ is network size, and randomly selecting a node $v$ to trigger with \textit{probability} $F_v$ at each step.  This approach is unsuitable when transmission rates are unbounded, since the time step $\Delta t$ cannot be rescaled \textit{a priori} to preserve the random node selection approach.  This is the case for the SI rule in our model. We define the transmission rate here to be proportional to edge state $\Fkm = \text{max} (p,\ \vm \cdot \boldsymbol{\lambda})$, where $\boldsymbol{\lambda} = \lambda \vw$, for a node in class $\vkvmclass$. Since $\vm$ and $\vlambda$ are unbounded in our simulations, as we impose no restriction on $\psi$ and $\Psi$, bursts of activity due to our renewal process model can result in arbitrarily high $F$. 

For this reason, we prefer an event-based Gillespie algorithm, which is equivalent, but advances in time by jumping to the next \textit{event}, rather than by uniform increments $\Delta t$. In the remainder of this section we discuss the implementation of sequences of these events.

\begin{algorithm}[H]
	\caption{Static network event sequence}\label{alg:static}
	\begin{algorithmic}[1]
		\Procedure{MonteCarloStatic}{$\Xi_{v}$}
		\While{$\Xi_v$ not empty}
		\State apply head of $\Xi_v$
		\EndWhile
		\EndProcedure
	\end{algorithmic}
\end{algorithm}

For a graph $\mathcal{G} (\mathcal{V}, \mathcal{E})$, we define two types of events, node events, defined over the node set $\mathcal{V}$, and edge events, defined over the edge set $\mathcal{E}$. A node event is implemented as the tuple $\{t_v, v\}$, and edge events $\{t_e, e_{uv}, s\}$, respectively. Node events amount to the infection of node $v$, at a time $t_v$, and edge events the positive or negative change in the state of edge $e_{uv}$, depending on the sign of the indicator $s = \pm 1$, according to our stochastic temporal network model. Monte Carlo simulation is implemented as two time ordered, dynamic sequences of events,  $\Xi_v$ and $\Xi_e$, for node and edge event types, respectively. While $\Xi_e$ is independent of node dynamics, both node end edge events feed back and cause a potential reordering of $\Xi_v$, as explained below. The node event sequence is initially of size $N = |\mathcal{V}|$, since we have one event for each node in the network, with the leading event being removed when all preceding edge activity has been carried out, i.e., when $t_v < t_e$, for the leading terms in each sequence. The action carried out is best illustrated by considering a static network, Algorithm \ref{alg:static}. Here, the \textit{apply head} instruction means to trigger the node in question, remove it from $\Xi_v$, and update the transmission rates of its neighbours, and in turn, their position in $\Xi_v$. In our model, the edge event sequences has approximately constant size, apart from some fluctuations due to the fact that $\eta / \langle \tau \rangle$ is only the expected edge state, the absolute number of events in $\eta$-memory can go up and down.  As networks increase in size, the size of the edge event sequence $\Xi_e$  converges to $\left(2 + \eta / \langle \tau \rangle \right)|\mathcal{E}|$.  Assuming that a non-zero level of noise is present, $p > 0$, then the $\Xi_v$ will eventually be emptied.

\begin{algorithm}[H]
	\caption{Temporal network event sequences}\label{alg:temporal}
	\begin{algorithmic}[1]
		\Procedure{MonteCarloTemporal}{$\Xi_{v}, \Xi_{e}$}
		\State $t_v, t_e\gets$ dequeue $\Xi_v$, $\Xi_e$
		\While{$\Xi_v$ not empty}
		\While{$t_{e} < t_{v}$}
		\State apply head of $\Xi_e$
		\State $t_v, t_e\gets$ dequeue $\Xi_v$, $\Xi_e$
		\EndWhile\label{euclidendwhile}
		\State apply head of $\Xi_v$
		\State $t_v\gets$ dequeue $\Xi_v$
		\EndWhile
		\EndProcedure
	\end{algorithmic}
\end{algorithm}

Algorithm \ref{alg:temporal} is the temporal extension of Algorithm \ref{alg:static}. In each, the time of infection of every node $v$ is determined at $t = 0$.  This is done by drawing from an exponential distribution with mean $F_v$, i.e., taking the natural logarithm of a uniform random variable on $(0,1)$, divided by $-F_v$. A node event is permanently erased from the sequence if $t_v < t_e$, as mentioned.  At this time, neighbours $u$ of $v$ have their transmission rates $F_u$ recalculated, as in Algorithm \ref{alg:static}, as they now have an additional infected neighbour. Additionally, the infection time of a node is recalculated if a change in its local neighbourhood takes place, such as activity on an adjacent edge, or the infection of a neighbouring node.  Since event sequences are time ordered, a node event is first erased from its position in the sequence, and then reinserted when such a calculation takes place. No such reordering take place for the edge event sequence, as once an event is inserted here, it is only removed when its time $t_e$ is at the front of the queue. While $t_e < t_v$, where the events in question are the leading events of each queue, there are two possible actions for the \textit{apply head} instruction for $\Xi_e$ in Algorithm \ref{alg:temporal}. The first, if $s = +1$, the leading edge event is erased and replaced with two new events on the same edge $e_{uv}$, occurring at time $t_e + \tau$, where $\tau$ is drawn from the interevent time distribution $\psi$. At the same time, an event is inserted for $t_e + \tau + \eta$, with $s = -1$, corresponding to the decrementing of that same edge $\eta$ time steps later. If an $s = -1$ edge event is leading the sequence, and we still have $t_e < t_v$, then the event is removed from the sequence without replacement. 

In the case of static networks, Gillespie algorithms allow massive speedup relative to the random selection approach. This is not the case for the node dynamics in our temporal network model, where a Gillespie event sequence affords very little speedup.  Here, the edge activity sequence is a bottleneck, since it still requires $|\mathcal{E}| dt / \langle \tau \rangle$ edge updates per $dt$ on average. As such, it is not for the benefits to speed that we use a Gillespie type event sequence for node updates, it is to account for arbitrarily large spikes in node transmission rates, particularly under the SI rule for temporal networks.

\subsection*{Inverse transform sampling of $\psi$ and $\Psi$}

In this section we discuss the numerical pipeline used to simulate renewal processes, a procedure that amounts to accurately sampling from the interevent time distributions $\psi(\tau)$ and $\Psi(\tau)$. Due to the large values of the standard deviation $\sigma_\tau$ to be examined in this work, currently available software could not be used to sample values of $\tau$.  While the excellent \texttt{<random>} library for C++ allows rapid sampling from the lognormal, Weibull and gamma distributions $\psi$, it appears to become inefficient for large $\sigma_\tau$, especially for the gamma distribution. In any case, directly sampling from $\Psi$, which often involves special functions, is beyond the scope of this library. Because of this, we build our own sampling routine.  

We favour an inverse transform sampling technique, where random values are sampled on an interval $(0,1)$, and evaluating the inverse of a cdf at this point provides a sample of the underlying pdf. That is, for a random variable $x \in (0,1)$, evaluating $\Psi^{-1} (x)$ provides a sample $\tau$ value from $\psi$. Further, random sampling of the \textit{residual} distribution requires finding the cdf of $\Psi$, and being able to approximate its inverse.

\begin{figure}
	\hspace*{0.5mm}
  \includegraphics[scale=1]{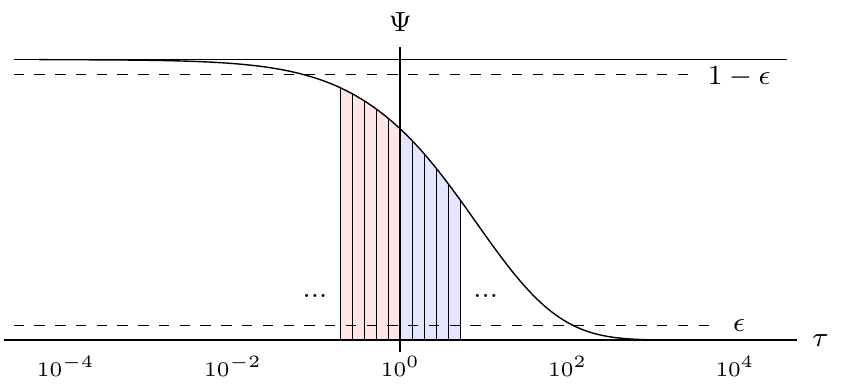}
	\caption{\textbf{Sampling probability distributions.} Numerical construction of $\Psi$, a probability density function that can be accurately evaluated at any point $\tau$, at some cost, for all distributions used in this work.  This grid results from iterating outwards from $\tau = \langle \tau \rangle = 1$, for increasing $\tau$ until $\Psi < \epsilon$, blue grid, and for decreasing $\tau$ until $\Psi > 1 - \epsilon$, red grid. Inverse transform sampling is then performed on the interval $(\epsilon, 1 - \epsilon)$ to provide samples of $\psi$, using a bisection method on the grid,   to find the corresponding $\tau$.  A third order spline interpolation on a logarithmic scale provides intermediate values of $\tau$.  Additionally, the grid can be cumulatively summed, and inverse transform sampling carried out on the spline of the resulting grid, to provide samples of $\Psi$.  \label{fig:fig12}}
\end{figure}

First, the problem of sampling from $\psi$ is that while $\Psi$ is known, its inverse generally is not. Second, we can't easily sample from $\Psi$ since \textit{its} cdf is generally unknown to begin with.  In fact, we often have enough difficulty simply evaluating $\Psi$, as is the case when $\psi$ is the gamma distribution, where $\Psi$ has no closed form. As a result, we can only determine $\Psi^{-1}$ approximately. We do this by first generating a grid of $\Psi$ values as shown in \sfref{fig:fig12}.  Due to the large $\sigma_\tau$ values of interest to us, it is difficult to estimate \textit{a priori} the desired upper and lower limits $\tau$ of the grid, which vary significantly depending on the choice of $\psi$. Said differently, we want $\epsilon$ to be as small as possible, in order to allow for the sampling of extreme values of $\tau$, which are critical to the dynamics of our system. Further, it goes without saying that taking the small $\epsilon$ limit ensures that $\langle \tau \rangle$ and $\sigma_\tau$ are respected, which is crucial since the latter is a control parameter in our experiment. As shown in \sfref{fig:fig12}, we iterate outward from $\tau = \langle \tau \rangle = 1$ uniformly on a logarithmic scale until a desired interval $(\epsilon, 1 - \epsilon)$ is obtained. An arbitrary precision library is required to accurately determine $\Psi$, and we prefer GNU MPFR, a multiple-precision binary floating-point library with correct rounding. See \SNprobability{} for details regarding the distributions themselves.

With such a grid accurately computed, we carry out third-order spline interpolation on a logarithmic scale to allow rapid evaluation at arbitrary points $\tau$ within the grid.  As a first application, the spline can be used to rapidly solve $\Psi (\tau) = x$ using a bisection technique.  This provides samples $\tau$ of $\psi$.  The grid can then be cumulatively summed to provide the cdf of $\Psi$ over the same domain. Performing approximate inverse transform sampling on the resulting grid returns samples $\tau$ of $\Psi$. 

It is worthwhile mentioning that rejection sampling techniques appear to be  out of the question here. This is due to the extreme bounding values of $\tau$ necessary for the standard deviations $\sigma_\tau$ studied in our choice of heavy-tailed distribution. Even for clever choices of enveloping functions for $\psi$ and $\Psi$, it is likely that the acceptance rates will be prohibitively low. Finding such functions remains an interesting challenge, but since our bisection approach converges exponentially quickly, it is hard to imagine that a choice of envelope exists that makes rejection sampling faster.

Finally, we comment on experiments involving very large standard deviations $\sigma_\tau$. Here, not even large networks running for a long time provide an unbiased sample of $\psi$. Following the central limit theorem, the standard deviation of the sampled mean interevent time equals $\sigma_\tau / \sqrt{n_s}$, if $n_s$ is the number of samples in question, which for the sake of argument is on the order of the number of edges in the network. High quality experimental results can be obtained by simulating large networks, or by averaging over realisations. Note further that large $\sigma_\tau$ experiments in our model happen to coincide with very long simulation times. As a result, even for large $\sigma_\tau$ there is little noise in our results due to the substantial runtime. Finally, $\xi_E$ plays a very important role, and amounts to the fraction of samples of $\Psi$ that are larger than $\eta$.  A consequence of this is that for large $\sigma_\tau$, most edges don't even participate, having drawn residual times at $t = 0$ that are longer than the duration of the experiment, determined by $\rho = 1 - e^{-pt}$. 

\section*{\SN~\arabic{suppnote}.~L\lowercase{aplace transform inversion}}
\refstepcounter{suppnote}

In the following sections we describe the numerical pipeline for obtaining the rates $\mu_j$ and $\nu_j$, as well as the edge-state distribution $E_j$. Although these values could be calculated by hand for the case of the exponential distribution, and maybe even the gamma distribution as its Laplace transform is known, we would like to be able to do this numerically for arbitrary $\psi (\tau)$, including those for which the Laplace transform of $\psi$ and $\Psi$ are not known.

These rates are expressed in terms of convolutions. Since we are interested in $j$-th order convolutions, for arbitrary positive integers $j$, we prefer to perform products in frequency space, as permitted by the Laplace transform.  A $j$-th order convolution for large $j$ increases exponentially in complexity, and can only be performed directly for very small $j$. Further, $E_j$ can be broad when $\sigma_\tau$ is large. The Laplace transform is defined as
\begin{equation}
  \hat{f}(s) = \mathcal{L}\{ f \} (s) = \int_{0}^{\infty}e^{-st}f(t)dt,
\end{equation}
where $f$ is a real-valued function of time $t$, and $\hat{f}$ a complex valued function of the complex variable $s = \sigma + i\omega$. For the Gaver-Stehfest algorithm in the following section, $s$ is always real, so we can set $\omega = 0$ and write 
\begin{equation}
  \hat{f}(s) = \int_{0}^{\infty}e^{-\sigma t}\cos\omega t f(t) dt - i\int_{0}^{\infty}e^{-\sigma t}\sin \omega tf(t)dt.
\end{equation}
This is necessary when the Laplace transform of $f$ is not know, and must be evaluated numerically, as is the case for the lognormal and Weibull distributions, used throughout this work. There exists a useful framework that unify these different algorithms~\cite{abate2006unified}, that express each approach in terms of a weighted sum of $\hat{f}$ values. We shall see that the computation of the forward Laplace transform is the bottleneck, as opposed to finding the weights. As such, the preferred method depends on how many numerical inversions of $f$ are required. This is $2M$ in the Gaver-Stehfest algorithm, $2M + 1$ in the Euler algorithm and $M$ in the Talbot algorithm, with $M$ controlling the desired accuracy (see~\cite{abate2006unified} for details). However, only the real integral need be found in the Gaver-Stehfest algorithm, which we prefer for this reason.

\subsection*{Gaver-Stehfest algorithm}\label{sec:gs}

We assume a Laplace transform $\hat{f}$ is known, and that we wish to recover the origin unknown function $f$. In our case, $f$ corresponds to convolutions of $\psi$ and $\Psi$. For any $t > 0$ and positive integer $M$, so-called Salzer summation yields the Gaver-Stehfest inversion formula,
\begin{equation}\label{eqn:gs_f}
  f(t, M) = \ln (2)t^{-1} \sum_{k = 1}^{2M}\zeta_{k}\hat{f} ( k \ln (2)t^{-1})
\end{equation}
where the weights $\zeta_{k}$ are given by
\begin{equation}\label{eqn:gs_zeta}
  \zeta_{k} = \dfrac{(-1)^{M + k}}{M!} \sum_{j = \left \lfloor{(k + 1) / 2}\right \rfloor}^{\min (k,M)} j^{M + 1} {M \choose j} {2j \choose j} {j \choose k - j}.
\end{equation}
Conveniently, the weights $\zeta_k$ are independent of the transform being inverted. This means that after a precision $M$ is chosen, weights can be stored in a vector $\zeta$ of dimension $2M$ to be reused for a number of transforms $\hat{f}$. We discuss the numerical implementation further in following sections, however mention here that we use the GNU Multiple Precision arithmetic library GMP for $\zeta_k$, in particular its integer summand, and the related MPFR library for floating point arithmetic for manipulating $\hat{f}$. Regarding the weights $\zeta_k$, a useful property for benchmarking is the fact that for all $M \geq 1$, 
\begin{equation}
  \sum_{k = 0}^{2M} \zeta_{k} = 0,
\end{equation}
due to the $(-1)^{M + k}$ factor in the definition of $\zeta_{k}$, these weights oscillate around zero. The Gaver-Stehfest algorithm requires a system precision of approximately $2.2M$ tracked bits, which we input to the constructors of variables in MPFR.

\subsection*{Numerical pipeline}

We now describe specifically how the inversion procedure in the previous section can be used to determine the values $E_j$, $\mu_j$ and $\nu_j$ described in preceding sections. For simplicity, we refer to these quantities in this section using vectors $E$, $\mu$ and $\nu$, whose dimension is determined by the size of the edge state space $n$ in our master equation, which need not be determined \textit{a priori}. Since $E_0$ and $\mu_0$ are special cases defined in the Methods section of the main text, we write $E = (E_1, E_2, \hdots, E_n)^T$ for the edge state distribution, and $\mu = (\mu_1, \mu_2, \hdots, \mu_n)^T$ and $\nu = (\nu_1, \nu_2, \hdots, \nu_n)^T$ for positive and negative edge transitions. The function $f(t, M)$ in the Gaver-Stehfest algorithm corresponds to $E_j (\eta, M)$, $\mu_j (\eta, M)$ and $\nu_j (\eta, M)$, where $M$ as before is a parameter of the Gaver-Stehfest algorithm that tunes the accuracy of the approximation.  


Calculating the distribution $E$, and the rate vectors $\mu$ and $\nu$ amounts to three separate matrix vector products.  We require the vector $\vzeta$, of dimension $2M$, as per the Gaver-Stehfest algorithm whose $k$-th element is given by \seref{eqn:gs_zeta}, and the Laplace transforms of $\psi$ and $\Psi$ evaluated at $s = k \ln (2) t^{-1}$, denoted $\psi_k$ and $\Psi_k$, as per \seref{eqn:Psi_hat}. We define $n$ as the maximum edge state, which can be as large as one likes here, it is limited only by $M$, which must be increased for increasing $n$.  These quantities are then used to populate the $n \times 2M$ dimensional matrices $\hat{E}$, $\hat{\mu}$ and $\hat{\nu}$, whose $jk$-th elements are given by
\begin{subequations}\label{eqn:hatEmunu}
	\begin{gather}
			[\hat{E}]_{jk} = \hat{\Psi}_k \cdot \hat{\psi}_k^{j - 1} \cdot \hat{\Psi}_k\\
		[\hat{\mu}]_{jk} = \hat{\Psi}_k \cdot \hat{\psi}_k^j\\
		[\hat{\nu]}_{jk} = \hat{\Psi}_k \cdot \hat{\psi}_k^{j - 1},
	\end{gather}
\end{subequations}
with $1 \leq j \leq n$ and $1 \leq k \leq 2M$. Determining the distribution of edge states $E$, and the edge transition rates $\mu$ and $\nu$, then amounts to the matrix-vector product of $\hat{E}$, $\hat{\mu}$ and $\hat{\nu}$, respectively, with $\vzeta$. As an illustration, the edge state distribution is given by
\begin{widetext}
\begin{equation}
	E = 
	\begin{pmatrix}
	  E_1 \\ E_2 \\ \vdots \\ E_n
	\end{pmatrix}
	=
    \setlength\arraycolsep{5pt}
	\begin{pmatrix}
		\hat{\Psi}_1 \cdot \hat{\psi}_1^0 \cdot \hat{\Psi}_1 & \hat{\Psi}_2 \cdot \hat{\psi}_2^0 \cdot \hat{\Psi}_2 & \hdots & \hat{\Psi}_{2M} \cdot \hat{\psi}_{2M}^0 \cdot \hat{\Psi}_{2M} \\
		\hat{\Psi}_1 \cdot \hat{\psi}_1^1 \cdot \hat{\Psi}_1 & \hat{\Psi}_2 \cdot \hat{\psi}_2^2 \cdot \hat{\Psi}_2 & \hdots & \hat{\Psi}_{2M} \cdot \hat{\psi}_{2M}^1 \cdot \hat{\Psi}_{2M} \\
		\vdots & \vdots & \ddots & \vdots \\
		\hat{\Psi}_1 \cdot \hat{\psi}_1^{n - 1} \cdot \hat{\Psi}_1 & \hat{\Psi}_2 \cdot \hat{\psi}_2^{n - 1} \cdot \hat{\Psi}_2 & \hdots & \hat{\Psi}_{2M} \cdot \hat{\psi}_{2M}^{n - 1} \cdot \hat{\Psi}_{2M} \\
	\end{pmatrix}
	\begin{pmatrix}
		\zeta_1 \\
		\zeta_2 \\
		\vdots \\
		\zeta_{2M}
	\end{pmatrix}.
\end{equation}
\end{widetext}
These quantities are calculated once at the start of the Runge-Kutta solution of the corresponding system, and don't change otherwise. Note that results have to be scaled by $\ln (2) t^{-1}$ as per the definition of $f(t, M)$ in the Gaver-Stehfest algorithm, and $\mu_j$ and $\nu_j$ normalised by $E_j$, to adhere to their definitions. The complete system to be solved is
\begin{subequations}
	\begin{align}
		  E &= \hat{E}   \vzeta\\
		\mu &= \hat{\mu} \vzeta\\
		\nu &= \hat{\nu} \vzeta.
	\end{align}
\end{subequations}
Examples of $\mu$ and $\nu$ values are provided in \sfref{fig:figS7} for the case of a gamma distribution $\psi$.

Implementation of the Gaver-Stehfest algorithm in C++ created confusion for some time, not because of the calculation of the weights $\zeta_k$, but in its product with $\hat{f}$. The authors of~\cite{abate2006unified} state that while it is clearly required for the weights $\zeta_k$, arbitrary precision is not required for handling $\hat{f}$. However, we find that double precision is not sufficient in general for $\hat{f}$, which can be confirmed by using double values in C++, or equivalently, setting a precision of 53 bits in a arbitrary precision library. We use a combination of the GMP library for the binomial coefficients, factorial, and exponential in \seref{eqn:gs_zeta}.  Division is performed after conversion to MPFR. Similarly, $\hat{f}$ in \seref{eqn:gs_f} is determined using MPFR, whether the Laplace transform is known in closed form, or if it is calculated from its integral definition.

\section*{\SN~\arabic{suppnote}.~D\lowercase{iffusion speed and noise}}
\refstepcounter{suppnote}

In the Methods section of the main text, we define the fraction of infections that are due to noise as $\rho_f$, and the spreading time relative to the pure-noise case as $t_f$. As shown in the inset of \MTmirror(a) of the main text, these quantities are very close to interchangeable. That is, any result expressed in terms of $t_f$ produces an almost identical picture in $\rho_f$, and vice versa.

\begin{figure}[b]
  \centering
  \hspace*{-1.5mm}
  \includegraphics[scale=1]{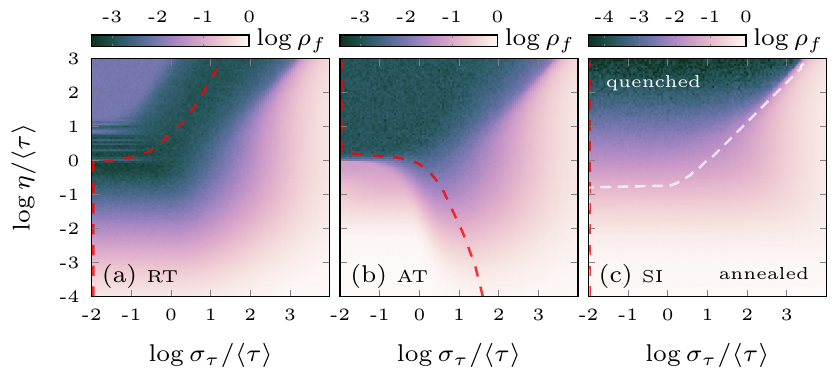}
  \caption{\textbf{Relative density heat maps.} Same experiment as for \MTheat{} in the main text, but plotting relative density of infections that are due to external noise, $\rho_f$, rather than the normalised spreading time, $t_f$. Results are almost identical, with the percolation transition preserved for both $\rho_f$ and $t_f$. See caption in \MTheat{} of the main text for simulation details. \label{fig:figS1}}
\end{figure}

As a concrete illustration we replot \MTheat{} of the main text in  terms of $\rho_f$, expressed there in terms of relative spreading time $t_f$. Results are shown in \sfref{fig:figS1}.  Upon inspection we note the landscape of $\rho_f$ is almost identical to that for $t_f$, albeit with slightly more variance in the relative noise measurement compared to spreading time, as seen in the reduced sharpness of the colours.  In particular the percolation transition, indicated by the dashed white line in \sfref{fig:figS1}(c), remains accurate. As expected, the slowest diffusion implies a complete dependence on external noise, as seen in the annealed regime of each plot where $\rho_f \approx 1$, meaning network topology plays a vanishingly small role in diffusion. On the other hand, rapid spreading occurs when network-induced infections are widespread, as in the quenched phase. This can be interpreted as external noise producing a strong catalytic effect, quantified by $1 / \rho_f$, the multiplicative effect of external noise. This give the number of total infections for every noise-induced infection, and is as large as $10^4$, as seen in the quenched regime of \sfref{fig:figS1}.


\begin{figure*}
  \hspace*{-1.5mm}
  \includegraphics[scale=1]{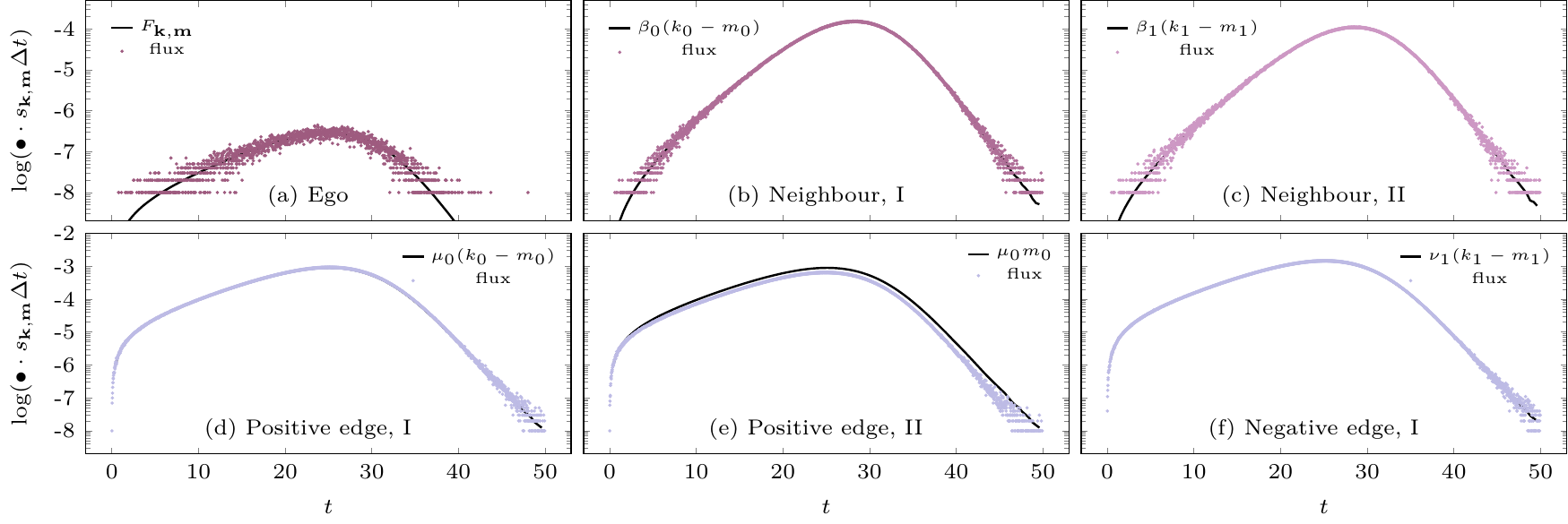}
  \caption{\textbf{Examination of mean field assumption for class transition rates.} Flux measurement for the set $\Svkvm$ with $\vk = (2,1)^T$ and $\vm = (1,0)^T$ during Monte Carlo simulation. Network size is $N = 10^8$, with a $3$-regular random degree distribution. Background noise is $p = 2 \times 10^{-4}$, with a relative threshold of $\phi = 0.4$. Interevent time distribution $\psi$ is power law, with $\alpha = 2.5$ and cutoffs $\tau = 1.01$ and $50$.  Memory duration is $\eta = 1$. Flux is measured over intervals of $\Delta t = 0.02$, shown as points, and compared to the density $\skm$ scaled by the indicated theoretical coefficients, shown as solid lines. \label{fig:figS5}}
\end{figure*}

\section*{\SN~\arabic{suppnote}.~M\lowercase{ean field approximation}}
\refstepcounter{suppnote}

In this Supplementary Note we examine the effectiveness of the mean field assumption of edge state transitions. The edge transition rates calculated in this work assume an infinitely large ensemble of stationary, uncorrelated renewal processes.  We expect a carefully implemented stochastic temporal network to agree precisely with theory, in the limit of large networks, which is what we confirm. In particular, by viewing the network as a single ensemble of edges, flux measurements indeed show theoretical results for $E_j$, $\mu_j$ and $\nu_j$ to be exact on a \textit{network wide} scale. However, our analysis involves partitioning our network into $2 \times |C|$ classes, namely, an infected and uninfected variant for each of the $|C|$ classes $\vkvmclass$ allowed by the system.  Clearly, the densities of nodes over these classes, and the transitions between them are highly dynamic, emptying and filling over the course of a spreading process. Further, transition rates are heterogeneous, particularly with respect to the transmission rate $\Fkm$ which varies from class to class.  The mean field approximation is to assume that edge transition rates for individual classes are the same as for a completely uncorrelated ensemble.

The reason that we expect differences in edge transition rates over these scales is as follows. The emergent rates $\mu_j$ and $\nu_j$ result from certain assumptions regarding  edge statistics at a microscopic level. This is formulated in terms of the \textit{history} distribution, or distribution of tuples $(\tau_1, \tau_2, \hdots, \tau_j)$ within each $\eta$ window, on each edge across a given set. This is illustrated in \sfref{fig:figS4}. Expressed in these terms, the mean field approach is to assume that the history distribution within a particular class $\vkvmclass$ is completely uncorrelated, and is equivalent to any randomly sampled subset of edges, or indeed the network edge set as a whole. To see how this assumption may break down, consider the flow of uninfected nodes through $C$ as a survival process, whereby a node enters an uninfected class $\vkvmclass$, only exiting and reentering an adjacent uninfected class if it does not become infected in the meantime. To further simplify this picture, consider $\vkvmclass$ to have $\Fkm = 1$, with neighbouring uninfected classes having $\Fkm = 0$, as may occur with a threshold model of infection.  In such a survival process, it is nodes with a history distribution $(\tau_1, \hdots, \tau_j)$ with short intervals in $\vkvmclass$, that are favourable to survival, exiting to adjacent uninfected classes. History distributions leading to long waiting times in $\vkvmclass$ are more likely to become infected, with survival times following an exponential distribution with mean $\Fkm$. As such, the history distribution of nodes exiting the class via infections are different to those that survive, and exit via edge or neighbour transitions.  Mechanisms like this, and the related effect due to neighbour transitions, gradually transform the history distribution of individual classes, resulting in deviations in the emergent edge transition rates $\mu$ and $\nu$. Although the assumption is broken on the scale of individual classes, it is of course preserved when taking all classes together.

We confirm this effect in an experiment whose results are shown in \sfref{fig:figS5}.  Here, we simulate a spreading process, and carefully record the densities of nodes in each class $\vkvmclass$ at all times $t$, as well the flows between classes over measurement windows of length $\Delta t = 0.02$. To allow $\Delta t$ to be as small as possible, and approximate the $dt$ in our analytics, we constrain the system to the smallest non-trivial configuration space $C$.  To this end, the degree distribution is $3$-regular random, and $\psi$ given by a power law with $\alpha = 2.5$, with lower and upper cutoffs of $\tau = 1.01$ and $50$, respectively.  By choosing $\eta = 1 < 1.01$, we ensure a two-level edge state space, where edges are in state $j = 0$ or $1$. The resulting configuration space has size $|C_3| = 20$, is connected, and resembles a smaller version of \sfref{fig:a1fig1}.  By setting network size as large as possible, here $N = 10^8$, the node set is diluted as little as possible over $C$.  We plot the flux measurements of the uninfected class with degree vectors $\vk = (2,1)^T$ and $\vm = (1,0)^T$. Node dynamics follow a relative threshold rule with $\phi = 0.4$, and background noise causing infection at a rate $p = 2 \times 10^{-4}$. As such, the transmission rate for the class in question is $\Fkm = p$.

Ego transition measurements are shown in \sfref{fig:figS5}(a). Since the class initially has density $\skm = 0$, with the network initialised to $\rho = 0$ at $t = 0$, initial measurements show only a handful of ego transitions per $\Delta t$ up to around $t = 10$, visible here thanks to the logscale. As expected, the rate of ego transition closely agrees with the measured value of $p = 2 \times 10^{-4}$, verified by scaling the total set density by $\Fkm \skm \Delta t$, given by the solid black curve. This transition is guaranteed to agree with theory if the Gillespie algorithm is correctly implemented, and serves as a useful benchmark in flux measurement experiments. Further, neighbour transition rates are verified in \sfref{fig:figS5}(b) and (c), for edges of type $j = 0$ and $1$ respectively. Time-dependent rates $\beta_j$ are calculated as per \seref{eqn:betasj}, using the set of $|C_3| = 20$ empirical densities $\skm$. Scaling $\skm$ for the class in question by $\beta_0 (k_0 - m_0)$ and $\beta_1 (k_1 - m_1)$ shows remarkable agreement with measured fluctuations.

Flux measurements of positive edge transitions for uninfected and infected neighbours are shown in \sfref{fig:figS5}(d) and (e), respectively. While agreement is excellent in (d), a clear deviation emerges in (e). Although the disagreement appears minor on a logarithmic scale, the theoretical $\mu_0$ is off by roughly $20\%$. The rate of negative edge transition in (f) is in good agreement with theory. A study of the complete state space shows that deviations as in (e) are common, but not systematic, with mean field estimates $\mu$ and $\nu$ sometimes overestimating, and sometimes underestimating the measured fluxes. The master equation solution of $\rho$ for this experiment, not shown here, miscalculated the overall spreading speed by about $15\%$. It is likely that the small configuration space here contributed to the error. In much larger systems, like in the main text, this effect is likely diluted, especially since $\mu$ and $\nu$ do not systematically over or underestimate the class-level edge transition rates. That is, a cancellation effect might emerge.

We confirm that the mean field assumption breaks down due to heterogeneities in transmission rates by studying a purely noise driven variant of the above experiments. That is, $\Fkm = p$ for all classes $\vkvmclass$. The same flux measurements, not shown here, are in perfect agreement with theory in such a setting. Further, the relative set sizes for constant $m$ rows of configuration space are exactly what one would expect given a degree distribution $p_k$, and a probability distribution $E_j$ that a randomly selected edge is in state $j$.

\section*{\SN~\arabic{suppnote}.~S\lowercase{kewness and entropy}}\label{sec:notes_probability}
\refstepcounter{suppnote}

\begin{figure*}
  \centering
  \includegraphics[scale=1]{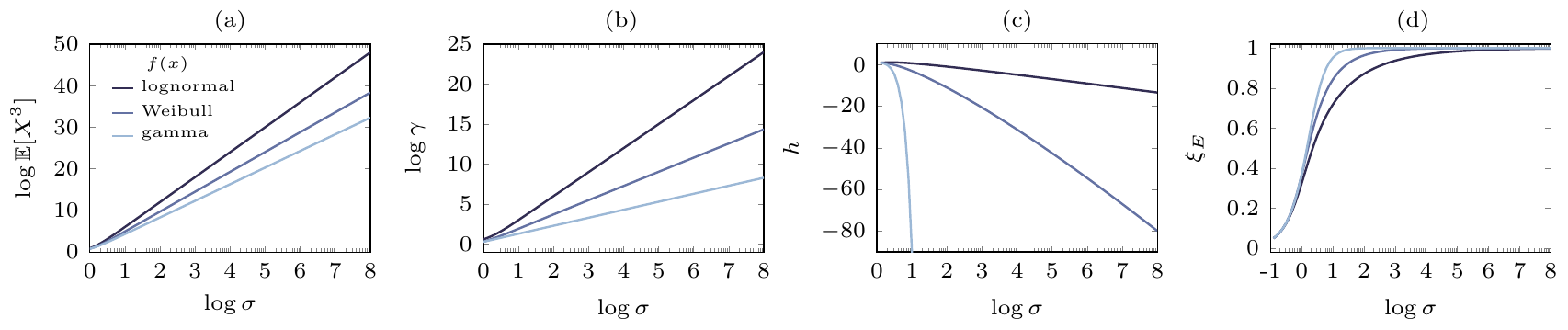}
  \caption{\textbf{Higher order moments of probability distributions.} Comparing skewness and entropy of probability distributions as a function of standard deviation. Mean is fixed to $\mu = 1$. (a) The third raw moment $\mathbb{E}[X^3]$, and  skewness $\gamma$ in (b), for the lognormal, Weibull and gamma distributions. The third raw moment is the only term that differs in the calculation of skewness. (c) Entropy follows the same trend as (a) and (b). We are interested in the connection between these quantities and $\xi_E$ in (d), defined in the main text, with $\eta = 1$.\label{fig:figS8}}
\end{figure*}

In this Supplementary Note we examine the skewness and differential entropy of the interevent time distributions used in this work, namely the lognormal, Weibull and gamma distributions. Our goal is to understand how these distributions differ after carefully controlling for their mean and standard deviation, $\mu$ and $\sigma$ respectively, and to this end propose skewness and entropy as measures that can be understood as ranking various distributions $\psi$ by the value of their effective sparsity $\xi_E$, discussed in the main text. 

We control for both the mean and standard deviation when comparing distributions in this work. For a random variable $X$, this is equivalent to controlling the first and second raw moments, $\mathbb{E}[X] = \mu$ and $\mathbb{E}[X^2] = \mu^2 + \sigma^2$.  We wish to examine the differences that remain between our chosen distributions after imposing these constraints.  To this end, it is logical to examine the \textit{third} raw moment $\mathbb{E}[X^3]$.  This is usually done by calculating the skewness $\gamma$, typically defined as the third standardised moment, or $\mathbb{E}[(X - \mu)^3]$ normalised by $\sigma^3$. The normalisation renders the moment scale invariant, meaning the information encoded in the skewness relates only to the ``shape'' of the distribution in question, and not its variance. Since we're interested in comparing skewness for constant mean and standard deviation, we use the expression
\begin{equation}\label{eqn:skewness}
  \gamma = \dfrac{\mathbb{E}[(X - \mu)^3]}{\sigma^3} = \dfrac{\mathbb{E}[X^3] - 3\mu\sigma^2 - \mu^3}{\sigma^3}.
\end{equation}
Clearly, skewness $\gamma$ differs from one distribution to another only in the third raw moment $\mathbb{E}[X^3]$, since we keep $\mu$ and $\sigma$ constant.  Plotting skewness $\gamma$ as a function of $\sigma$ for constant $\mu = 1$ leads to the plot in \sfref{fig:figS8}, for the lognormal, Weibull and gamma distributions.  The large differences in skewness here correspond to observations regarding effective sparsity $\xi_E$ in earlier parts of this work. As such, the skewness of a distribution provides a good rule-of-thumb for comparing $\xi_E$ for different distributions.

In addition, we calculate the differential, or information entropy for each distribution, defined as 
\begin{equation}
  h = -\int_0^\infty f(x) \ln f(x) dx,
\end{equation}
for distributions defined for $x \in (0,\infty )$, as is the case for the lognormal, Weibull and gamma distribution. The motivation for considering the differential entropy is the observation that it was the lognormal distribution, the maximum entropy distribution for which the mean and variance of $\ln (X)$ are specified, that produced the fastest diffusion times in the main text.

As shown in \sfref{fig:figS8}, the relative values of skewness and entropy agree qualitatively with expectation. Simply put, the relative values of skewness $\gamma$, shown in plot (b), and differential entropy $h$, shown in plot (c), agree qualitatively with the relative values of effective sparsity $\xi_E$, shown in plot (d). This supports the rule-of-thumb that the greater the skewness and entropy, the lower the effective sparsity. 

\section*{\SN~\arabic{suppnote}.~P\lowercase{robability distributions}}\label{sec:notes_probability}
\refstepcounter{suppnote}

In this Supplementary Note we detail the probability distributions used in this work, in particular those used for the interevent time distribution $\psi$, and its tail distribution $\Psi$. For simplicity, the notation used in this Supplementary Note is entirely self contained, and we associate pdfs $f(x)$ with $\psi(\tau)$, and their cdfs $F(x)$ with $1 - \Psi (\tau)$. 

\begin{table*}[t]
  \caption{\textbf{Properties of probability distributions.} Properties of the two-parameter probability distributions used to model interevent time in this work. For each we note the probability density function $f_X(x)$, the cumulative density function $F_X(x)$, the mean $\mu$ and variance $\sigma^2$, as well as the third moment $\mathbb{E}[X^3]$ and skewness $\gamma$, and differential entropy $h$.
  \label{tab:tab1}}
	\begin{ruledtabular}
	    \small
	    \renewcommand{\arraystretch}{2.3}
		\begin{tabular}{c c c c}
	    & lognormal$^\dagger$ & Weibull$^\ddagger$ & gamma \\
	  	\hline 
	    $f_X(x)$ 
	    & 
	    $\dfrac{1}{x \sqrt{2\pi \tilde{\sigma}^2}} \exp \left[ -\dfrac{\left(\ln x - \tilde{\mu}\right)^2}{\sqrt{2\pi \tilde{\sigma}^2}} \right]$
	    &
	    $\dfrac{k}{\lambda}\left(\dfrac{x}{\lambda}\right)^{k - 1}e^{-(x / \lambda)^{k}}$
	    &
	    $\dfrac{1}{\Gamma(\alpha)\beta^{\alpha}}x^{\alpha - 1}e^{-\tfrac{x}{\beta}}$
	    \\
	    $F_X(x)$
	    &
	    $\dfrac{1}{2} + \dfrac{1}{2}\text{erf}\left(\dfrac{\ln x - \tilde{\mu}}{\sqrt{2\tilde{\sigma}^2}} \right)$
	    &
	    $1 - e^{-(x / \lambda)^{k}}$
	    &
	    $\dfrac{1}{\Gamma (\alpha)}\gamma (\alpha, \tfrac{x}{\beta})$
	    \\
	    $\mu$
	    &
	    $\ln \left( \dfrac{\mu^{2}}{\sqrt{\mu^{2} + \sigma^2}} \right)$
	    &
	    $\lambda \Gamma (1 + 1/k)$
	    &
	    $\alpha \beta$
	    \\
	    $\sigma^2$
	    &
	    $\ln \left(\dfrac{\mu^2 + \sigma^2}{\mu^2}\right)$
	    &
	    $\lambda^2 \left[ \Gamma (1 + 2/k) - \Gamma^2 (1 + 1/k) \right]$
	    &
	    $\alpha \beta^2$
	    \\
	    $\mathbb{E}[X^3]$
	    &
	    $\dfrac{\left(\mu^2 + \sigma^2 \right)^3}{\mu^3}$
	    &
	    $\lambda^3 \Gamma (1 + 3 / k)$
	    &
	    $\mu^3 + 3\mu\sigma^2 + \dfrac{2\sigma^4}{\mu}$
	    \\
	    $\gamma$
	    &
	    $\dfrac{3\mu^2\sigma + \sigma^3}{\mu^3}$
	    &
	    -
	    &
	    $\dfrac{2\sigma}{\mu}$
	    \\
	    $h$
	    &
	    $\tilde{\mu} + \ln (2\pi e \tilde{\sigma}^2)$
	    &
	    $\left(1 - 1/k \right)\gamma_e + \ln \left(\lambda / k\right) + 1$
	    &
	    $\alpha + \ln (\alpha \Gamma (\alpha)) + (1 - \alpha)\psi(\alpha)$
		\end{tabular}
		\\[2pt]
        {\footnotesize $\dagger$ Note that the provided mean and variance correspond to the underlying normal distribution with mean $\tilde{\mu}$ and variance $\tilde{\sigma}^2$}\\[-2pt]
        {\footnotesize $\ddagger$ We omit skewness since it doesn't simplify like the other distributions. Further, $\gamma_e$ is the Euler-Mascheroni constant.}
	\end{ruledtabular}
\end{table*}	


\textit{Lognormal distribution}. The lognormal distribution is defined with respect to an underlying normal distribution with mean $\tilde{\mu}$ and variance $\tilde{\sigma}^2$. These are related to the lognormal mean and variance $\mu$ and $\sigma^2$ by the relations given in \stref{tab:tab1}, and can be straightforwardly inverted. Provided $\mu$ and $\sigma$, it has the largest skewness $\gamma$ of all distributions studied here, as well as the largest differential entropy $h$.  This is not surprising, given that the lognormal can be derived using maximum entropy principles, as discussed in the previous section, and illustrated in \sfref{fig:figS8}.


\textit{Gamma distribution}. The gamma distribution is related to the gamma function, $\Gamma (s)$, described below. The expressions for its mean and variance  \stref{tab:tab1} can be easily inverted, resulting in expressions for $\alpha$ and $\beta$ the provide the desired moments. Note that we set $\mu = 1$ for the temporal network experiments in this work, meaning that when the \textit{shape} parameter $\alpha = \beta = 1$, coinciding with $\sigma = 1$, we recover the exponential distribution. For values $\alpha < 1$, meaning $\sigma > 1$, the qualitative shape of the exponential is maintained, with an increasingly heavy tail. In contrast, when $\alpha > 1$, meaning $\sigma < 1$, the shape changes, and like the lognormal, $f$ goes to zero in the small $x$ limit. For large $\alpha$, meaning small $\sigma$, the gamma distribution tends to the Dirac delta function. In \sfref{fig:figS8} we observe that the gamma distribution is the least skewed of the distributions considered here, for comparable $\mu$ and $\sigma$. Entropy is given using $\psi$, the digamma function, defined in the following section under the Weibull distribution.

\textit{Variants of the gamma function}. We describe here the necessary approximations in order to numerically construct the cdf of the gamma distribution.  Unfortunately, the cdf as stated here is circular, with no closed form expression available. In this section, we discuss a number of series expansions that are necessary to numerically evaluate $\gamma (\alpha, \tfrac{x}{\beta})$, the lower incomplete gamma function. The gamma function $\Gamma (s)$ is defined, and related to its incomplete variants, as
\begin{subequations}
  \begin{eqnarray}
    \Gamma (s) &=& \int_{0}^{\infty}t^{s - 1}e^{-t}dt\\
               &=& \int^{x}_{0}t^{s - 1}e^{-t}dt + \int^{\infty}_{x}t^{s - 1}e^{-t}dt\\
               &=& \gamma (s, x) + \Gamma (s, x).
  \end{eqnarray}
\end{subequations}
In other words, the upper and lower incomplete gamma functions are defined by partitioning the integral according to $(0, \infty) = (0, x] \cup [x, \infty)$, given by $\gamma (s,x)$ and $\Gamma (s, x)$, respectively.  Sampling from the lower incomplete gamma function will be crucial when initialising our temporal network system at steady state. The choice of series approximation for the lower-incomplete gamma function depends on the shape parameter $\alpha$, and in turn the standard deviation $\sigma$. For small $\alpha$, meaning large $\sigma$, we use
\begin{equation}
  \gamma (s, x) = e^{-x}x^{s}\Gamma (s) \sum_{n = 0}^{\infty}\dfrac{x^{n}}{\Gamma (s + n + 1)}.
\end{equation}
This well known approximation, although efficient for large values of standard deviation, becomes prohibitively slow for very large values of $\alpha$, meaning very small values of $\sigma$. At this scale, specifically when $\alpha > 1$ and $\sigma < 1$, we approximate the lower-incomplete gamma function as 
\begin{equation}
  \gamma (s, x) = e^{-x}x^{s} \sum_{n = 0}^{\infty}\dfrac{x^n}{s^{\overline{n + 1}}}
\end{equation}
which can be computed recursively using a small number of multiplication and division operations at each step. Unfortunately this approximation fails for large $\sigma$, and must be only be used in the small $\sigma$, which we do purely for efficiency, since the preceding approximation converges everywhere.  Here, $s^{\overline{n + 1}}$ is the Pochhammer symbol. 



\textit{Weibull distribution}. Closely related to the gamma distribution is the Weibull distribution. As can be seen in \stref{tab:tab1}, the Weibull distribution cannot be easily parameterised by its mean and standard deviation, as was the case for the lognormal and gamma distribution. The tuple $(k, \lambda)$ providing desired mean and standard deviation can be found using gradient descent. Note that like the gamma distribution, when $k = \lambda = 1$, we recover the exponential distribution. Like $\alpha$ in the gamma distribution, $k$ controls the shape, with large $k$ tending towards the Dirac delta function, and small $k$ an increasingly right skewed, heavy tailed form.


\textit{Tuning parameters using gradient descent}. We could not find a reference to address the problem of parameterising the Weibull distribution by its mean $\mu$ and standard deviation $\sigma$. Not finding an existing solution to this problem, we find the values of $k$ and $\lambda$ giving the desired $\mu$ and $\sigma$ using gradient descent. In the following calculations, we actually use the variance denoted by $\nu = \sigma^2$ for simplicity, given the form of \seref{eq:appf_sigmaweibull}. We find $k$ and $\lambda$ by locating the minimum of the loss surface defined by the function
\begin{equation}
  l(\mu_i, \nu_i) = \left(\dfrac{\mu_i - \mu}{\mu} \right)^2 + \left(\dfrac{\nu_i - \nu}{\nu}\right)^2,
\end{equation}
where $\mu_i = \mu_i (k_i, \lambda_i)$ and $\nu_i = \nu_i (k_i, \lambda_i)$, as per \stref{tab:tab1}, are the values of the mean and variance at the $i$-th step of the procedure, and are continuous variables here.  The values $\mu$ and $\nu$ are the target, and are considered constant in the following.  As such, $\mu_i$ and $\nu_i$ are continuous variables. Due to the nature of the experiments in the main text, it is crucial to normalise the relative error in each term, by $\mu$ and $\nu$ respectively. This is because $\nu$ varies over orders of magnitude, while $\mu$ remains fixed. For the same reason, the displacement at each step of the algorithm is determined on logarithmic scales. We perform this optimisation once, and store the resulting $(\mu, \nu, k, \lambda)$ tuple in a table for reuse. 
\begin{widetext}
\noindent The gradient of $l(\mu_i, \nu_i)$ is
\begin{equation}
  \nabla l(\mu_i, \nu_i) = \partial_k l(\mu_i, \nu_i) \boldsymbol{\hat{k}} + \partial_{\lambda} l(\mu_i, \nu_i) \boldsymbol{\hat{\lambda}},
\end{equation}
with $k$ and $\lambda$ components given by
\begin{equation}
		\partial_k l(\mu_i, \nu_i) = 2\left(\dfrac{\mu_i - \mu}{\mu}\right)\dfrac{1}{\mu}\partial_k \mu_i + 2\left(\dfrac{\nu_i - \nu}{\nu}\right)\dfrac{1}{\nu}\partial_k \nu_i
\end{equation}
and
\begin{equation}
		\partial_{\lambda} l(\mu_i, \nu_i) = 2\left(\dfrac{\mu_i - \mu}{\mu}\right)\dfrac{1}{\mu}\partial_\lambda \mu_i + 2\left(\dfrac{\nu_i - \nu}{\nu}\right)\dfrac{1}{\nu}\partial_\lambda \nu_i,
\end{equation}
with partial derivatives
\begin{gather}
  \partial_\lambda \mu_i = \Gamma(1 + 1/k),\\
  \partial_k \mu_i = -\dfrac{\lambda}{k^2}\Gamma(1 + 1/k)\psi(1 + 1/k),
\end{gather}
and
\begin{gather}
  \partial_\lambda \nu_i = 2\lambda \left[ \Gamma(1 + 2 / k) - \Gamma^2(1 + 1/k) \right],\\
  \partial_k \nu_i = -\dfrac{2\lambda^2}{k^2}\left[\Gamma(1 + 2 / k)\psi(1 + 2 / k) - \Gamma^2(1 + 1 / k)\psi (1 + 1/k)\right],
\end{gather}
\end{widetext}
respectively. Here $\psi$ is the so-called digamma function, which was previously required for the calculation of the entropy of the gamma distribution. It is defined as the logarithmic derivative of the gamma function, or 
\begin{equation}
  \psi (z) = \dfrac{d}{dz}\ln (\Gamma (z)) = \dfrac{\Gamma^{\prime} (z)}{\Gamma (z)}.
\end{equation}
It can be found using a simple series approximation, 
\begin{equation}
  \psi (z) = -\gamma + \sum_{n = 0}^{\infty}\left(\dfrac{1}{n + 1} - \dfrac{1}{n + z}\right),
\end{equation}
for $z \neq -1, -2, -3, \hdots$, where $\gamma$ here denotes the Euler-Mascheroni constant. Although the series is infinite, in practice it converges quite rapidly, even when applying strict thresholds. 
Along with an inputted step size, this provides the next point $(\mu_i, \nu_i, k_i, \lambda_i)$ in the procedure. Adaptive step size is incorporated in gradient descent, where step size $h_i = l_i$ if $l_i < 0.1$, and $h_i = 0.1$ otherwise. The gradient is estimated manually by calculating the local neighbourhood of the current point $k_i$, $\lambda_i$. Stochastic gradient descent techniques may be incorporated to improve the convergence rate.

Note that a multiple precision library appears to be necessary even when the desired precision in the loss function $l$ not beyond the bounds of default machine precision, such as \texttt{double} and \texttt{long double}. As always, this induces a cost in terms of computation time, so we calculate all the tuples $(\mu, \sigma, k, \lambda)$ that are needed for the entire calculation just once, and then store these in a lookup table.


%% file: manuscript_pilot.bbl
\begin{thebibliography}{10}
\expandafter\ifx\csname url\endcsname\relax
  \def\url#1{\texttt{#1}}\fi
\expandafter\ifx\csname urlprefix\endcsname\relax\def\urlprefix{URL }\fi
\providecommand{\bibinfo}[2]{#2}
\providecommand{\eprint}[2][]{\url{#2}}

\bibitem{holme2012temporal}
\bibinfo{author}{Holme, P.} \& \bibinfo{author}{Saram{\"a}ki, J.}
\newblock \bibinfo{title}{Temporal networks}.
\newblock \emph{\bibinfo{journal}{Phys. Rep.}} \textbf{\bibinfo{volume}{519}},
  \bibinfo{pages}{97--125} (\bibinfo{year}{2012}).

\bibitem{masuda2016guide}
\bibinfo{author}{Masuda, N.} \& \bibinfo{author}{Lambiotte, R.}
\newblock \emph{\bibinfo{title}{A Guide to Temporal Networks}}
  (\bibinfo{publisher}{World Scientific}, \bibinfo{year}{2016}).

\bibitem{holme2015modern}
\bibinfo{author}{Holme, P.}
\newblock \bibinfo{title}{Modern temporal network theory: a colloquium}.
\newblock \emph{\bibinfo{journal}{Eur. Phys. J. B}}
  \textbf{\bibinfo{volume}{88}}, \bibinfo{pages}{234} (\bibinfo{year}{2015}).

\bibitem{barabasi2005origin}
\bibinfo{author}{Barab{\'a}si, A.-L.}
\newblock \bibinfo{title}{The origin of bursts and heavy tails in human
  dynamics}.
\newblock \emph{\bibinfo{journal}{Nature}} \textbf{\bibinfo{volume}{435}},
  \bibinfo{pages}{207} (\bibinfo{year}{2005}).

\bibitem{goh2008burstiness}
\bibinfo{author}{Goh, K.-I.} \& \bibinfo{author}{Barab{\'a}si, A.-L.}
\newblock \bibinfo{title}{Burstiness and memory in complex systems}.
\newblock \emph{\bibinfo{journal}{EPL}} \textbf{\bibinfo{volume}{81}},
  \bibinfo{pages}{48002} (\bibinfo{year}{2008}).

\bibitem{karsai2012universal}
\bibinfo{author}{Karsai, M.}, \bibinfo{author}{Kaski, K.},
  \bibinfo{author}{Barab{\'a}si, A.-L.} \& \bibinfo{author}{Kert{\'e}sz, J.}
\newblock \bibinfo{title}{Universal features of correlated bursty behaviour}.
\newblock \emph{\bibinfo{journal}{Sci. Rep.}} \textbf{\bibinfo{volume}{2}},
  \bibinfo{pages}{1--7} (\bibinfo{year}{2012}).

\bibitem{davidsen2013earthquake}
\bibinfo{author}{Davidsen, J.} \& \bibinfo{author}{Kwiatek, G.}
\newblock \bibinfo{title}{Earthquake interevent time distribution for induced
  micro-, nano-, and picoseismicity}.
\newblock \emph{\bibinfo{journal}{Phys. Rev. Lett.}}
  \textbf{\bibinfo{volume}{110}}, \bibinfo{pages}{068501}
  (\bibinfo{year}{2013}).

\bibitem{deArcangelis2006universality}
\bibinfo{author}{de~Arcangelis, L.}, \bibinfo{author}{Godano, C.},
  \bibinfo{author}{Lippiello, E.} \& \bibinfo{author}{Nicodemi, M.}
\newblock \bibinfo{title}{Universality in solar flare and earthquake
  occurrence}.
\newblock \emph{\bibinfo{journal}{Phys. Rev. Lett.}}
  \textbf{\bibinfo{volume}{96}}, \bibinfo{pages}{051102}
  (\bibinfo{year}{2006}).

\bibitem{turnbull2005string}
\bibinfo{author}{Turnbull, L.}, \bibinfo{author}{Dian, E.} \&
  \bibinfo{author}{Gross, G.}
\newblock \bibinfo{title}{The string method of burst identification in neuronal
  spike trains}.
\newblock \emph{\bibinfo{journal}{J. Neurosci. Methods}}
  \textbf{\bibinfo{volume}{145}}, \bibinfo{pages}{23--35}
  (\bibinfo{year}{2005}).

\bibitem{karsai2018bursty}
\bibinfo{author}{Karsai, M.}, \bibinfo{author}{Jo, H.-H.} \&
  \bibinfo{author}{Kaski, K.}
\newblock \emph{\bibinfo{title}{Bursty Human Dynamics}}
  (\bibinfo{publisher}{Springer}, \bibinfo{year}{2018}).

\bibitem{karsai2011small}
\bibinfo{author}{Karsai, M.} \emph{et~al.}
\newblock \bibinfo{title}{Small but slow world: how network topology and
  burstiness slow down spreading}.
\newblock \emph{\bibinfo{journal}{Phys. Rev. E}} \textbf{\bibinfo{volume}{83}},
  \bibinfo{pages}{025102} (\bibinfo{year}{2011}).

\bibitem{lambiotte2013burstiness}
\bibinfo{author}{Lambiotte, R.}, \bibinfo{author}{Tabourier, L.} \&
  \bibinfo{author}{Delvenne, J.-C.}
\newblock \bibinfo{title}{Burstiness and spreading on temporal networks}.
\newblock \emph{\bibinfo{journal}{Eur. Phys. J. B}}
  \textbf{\bibinfo{volume}{86}}, \bibinfo{pages}{320} (\bibinfo{year}{2013}).

\bibitem{jo2014analytically}
\bibinfo{author}{Jo, H.-H.}, \bibinfo{author}{Perotti, J.~I.},
  \bibinfo{author}{Kaski, K.} \& \bibinfo{author}{Kert{\'e}sz, J.}
\newblock \bibinfo{title}{Analytically solvable model of spreading dynamics
  with non-{P}oissonian processes}.
\newblock \emph{\bibinfo{journal}{Phys. Rev. X}} \textbf{\bibinfo{volume}{4}},
  \bibinfo{pages}{011041} (\bibinfo{year}{2014}).

\bibitem{horvath2014spreading}
\bibinfo{author}{Horv{\'a}th, D.~X.} \& \bibinfo{author}{Kert{\'e}sz, J.}
\newblock \bibinfo{title}{Spreading dynamics on networks: the role of
  burstiness, topology and non-stationarity}.
\newblock \emph{\bibinfo{journal}{New J. Phys.}} \textbf{\bibinfo{volume}{16}},
  \bibinfo{pages}{073037} (\bibinfo{year}{2014}).

\bibitem{williams2019effects}
\bibinfo{author}{Williams, O.~E.}, \bibinfo{author}{Lillo, F.} \&
  \bibinfo{author}{Latora, V.}
\newblock \bibinfo{title}{Effects of memory on spreading processes in
  non-{M}arkovian temporal networks}.
\newblock \emph{\bibinfo{journal}{New J. Phys.}} \textbf{\bibinfo{volume}{21}},
  \bibinfo{pages}{043028} (\bibinfo{year}{2019}).

\bibitem{vazquez2007impact}
\bibinfo{author}{Vazquez, A.}, \bibinfo{author}{R\'acz, B.},
  \bibinfo{author}{Luk\'acs, A.} \& \bibinfo{author}{Barab\'asi, A.-L.}
\newblock \bibinfo{title}{Impact of non-{P}oissonian activity patterns on
  spreading processes}.
\newblock \emph{\bibinfo{journal}{Phys. Rev. Lett.}}
  \textbf{\bibinfo{volume}{98}}, \bibinfo{pages}{158702}
  (\bibinfo{year}{2007}).

\bibitem{mancastroppa2019burstiness}
\bibinfo{author}{Mancastroppa, M.}, \bibinfo{author}{Vezzani, A.},
  \bibinfo{author}{Mu{\~{n}}oz, M.~A.} \& \bibinfo{author}{Burioni, R.}
\newblock \bibinfo{title}{Burstiness in activity-driven networks and the
  epidemic threshold}.
\newblock \emph{\bibinfo{journal}{J. Stat. Mech.: Theory Exp.}}
  \textbf{\bibinfo{volume}{2019}}, \bibinfo{pages}{053502}
  (\bibinfo{year}{2019}).

\bibitem{pastorsatorras2015epidemic}
\bibinfo{author}{Pastor-Satorras, R.}, \bibinfo{author}{Castellano, C.},
  \bibinfo{author}{Van~Mieghem, P.} \& \bibinfo{author}{Vespignani, A.}
\newblock \bibinfo{title}{Epidemic processes in complex networks}.
\newblock \emph{\bibinfo{journal}{Rev. Mod. Phys.}}
  \textbf{\bibinfo{volume}{87}}, \bibinfo{pages}{925--979}
  (\bibinfo{year}{2015}).

\bibitem{vespignani2020modelling}
\bibinfo{author}{Vespignani, A.} \emph{et~al.}
\newblock \bibinfo{title}{Modelling {COVID}-19}.
\newblock \emph{\bibinfo{journal}{Nat. Rev. Phys.}} \bibinfo{pages}{1--3}
  (\bibinfo{year}{2020}).

\bibitem{starnini2017equivalence}
\bibinfo{author}{Starnini, M.}, \bibinfo{author}{Gleeson, J.~P.} \&
  \bibinfo{author}{Bogu{\~n}{\'a}, M.}
\newblock \bibinfo{title}{Equivalence between non-{M}arkovian and {M}arkovian
  dynamics in epidemic spreading processes}.
\newblock \emph{\bibinfo{journal}{Phys. Rev. Lett.}}
  \textbf{\bibinfo{volume}{118}}, \bibinfo{pages}{128301}
  (\bibinfo{year}{2017}).

\bibitem{liu2014controlling}
\bibinfo{author}{Liu, S.}, \bibinfo{author}{Perra, N.},
  \bibinfo{author}{Karsai, M.} \& \bibinfo{author}{Vespignani, A.}
\newblock \bibinfo{title}{Controlling contagion processes in activity driven
  networks}.
\newblock \emph{\bibinfo{journal}{Phys. Rev. Lett.}}
  \textbf{\bibinfo{volume}{112}}, \bibinfo{pages}{118702}
  (\bibinfo{year}{2014}).

\bibitem{masuda2017temporal}
\bibinfo{author}{Masuda, N.} \& \bibinfo{author}{Holme, P.}
\newblock \emph{\bibinfo{title}{Temporal Network Epidemiology}}
  (\bibinfo{publisher}{Springer}, \bibinfo{year}{2017}).

\bibitem{masuda2020small}
\bibinfo{author}{Masuda, N.} \& \bibinfo{author}{Holme, P.}
\newblock \bibinfo{title}{Small inter-event times govern epidemic spreading on
  networks}.
\newblock \emph{\bibinfo{journal}{Phys. Rev. Res.}}
  \textbf{\bibinfo{volume}{2}}, \bibinfo{pages}{023163} (\bibinfo{year}{2020}).

\bibitem{miritello2011dynamical}
\bibinfo{author}{Miritello, G.}, \bibinfo{author}{Moro, E.} \&
  \bibinfo{author}{Lara, R.}
\newblock \bibinfo{title}{Dynamical strength of social ties in information
  spreading}.
\newblock \emph{\bibinfo{journal}{Phys. Rev. E}} \textbf{\bibinfo{volume}{83}},
  \bibinfo{pages}{045102} (\bibinfo{year}{2011}).

\bibitem{hiraoka2018correlated}
\bibinfo{author}{Hiraoka, T.} \& \bibinfo{author}{Jo, H.-H.}
\newblock \bibinfo{title}{Correlated bursts in temporal networks slow down
  spreading}.
\newblock \emph{\bibinfo{journal}{Sci. Rep.}} \textbf{\bibinfo{volume}{8}},
  \bibinfo{pages}{1--12} (\bibinfo{year}{2018}).

\bibitem{min2011spreading}
\bibinfo{author}{Min, B.}, \bibinfo{author}{Goh, K.-I.} \&
  \bibinfo{author}{Vazquez, A.}
\newblock \bibinfo{title}{Spreading dynamics following bursty human activity
  patterns}.
\newblock \emph{\bibinfo{journal}{Phys. Rev. E}} \textbf{\bibinfo{volume}{83}},
  \bibinfo{pages}{036102} (\bibinfo{year}{2011}).

\bibitem{rocha2011simulated}
\bibinfo{author}{Rocha, L.~E.}, \bibinfo{author}{Liljeros, F.} \&
  \bibinfo{author}{Holme, P.}
\newblock \bibinfo{title}{Simulated epidemics in an empirical spatiotemporal
  network of 50,185 sexual contacts}.
\newblock \emph{\bibinfo{journal}{PLoS Comput. Biol.}}
  \textbf{\bibinfo{volume}{7}}, \bibinfo{pages}{e1001109}
  (\bibinfo{year}{2011}).

\bibitem{granovetter1978threshold}
\bibinfo{author}{Granovetter, M.}
\newblock \bibinfo{title}{Threshold models of collective behavior}.
\newblock \emph{\bibinfo{journal}{Am. J. Sociol.}}
  \textbf{\bibinfo{volume}{83}}, \bibinfo{pages}{1420--1443}
  (\bibinfo{year}{1978}).

\bibitem{karsai2016local}
\bibinfo{author}{Karsai, M.}, \bibinfo{author}{I{\~n}iguez, G.},
  \bibinfo{author}{Kikas, R.}, \bibinfo{author}{Kaski, K.} \&
  \bibinfo{author}{Kert{\'e}sz, J.}
\newblock \bibinfo{title}{Local cascades induced global contagion: how
  heterogeneous thresholds, exogenous effects, and unconcerned behaviour govern
  online adoption spreading}.
\newblock \emph{\bibinfo{journal}{Sci. Rep.}} \textbf{\bibinfo{volume}{6}},
  \bibinfo{pages}{27178} (\bibinfo{year}{2016}).

\bibitem{watts2002simple}
\bibinfo{author}{Watts, D.~J.}
\newblock \bibinfo{title}{A simple model of global cascades on random
  networks}.
\newblock \emph{\bibinfo{journal}{Proc. Natl. Acad. Sci. U.S.A.}}
  \textbf{\bibinfo{volume}{99}}, \bibinfo{pages}{5766--5771}
  (\bibinfo{year}{2002}).

\bibitem{gleeson2008cascades}
\bibinfo{author}{Gleeson, J.~P.}
\newblock \bibinfo{title}{Cascades on correlated and modular random networks}.
\newblock \emph{\bibinfo{journal}{Phys. Rev. E}} \textbf{\bibinfo{volume}{77}},
  \bibinfo{pages}{046117} (\bibinfo{year}{2008}).

\bibitem{unicomb2018threshold}
\bibinfo{author}{Unicomb, S.}, \bibinfo{author}{I{\~n}iguez, G.} \&
  \bibinfo{author}{Karsai, M.}
\newblock \bibinfo{title}{Threshold driven contagion on weighted networks}.
\newblock \emph{\bibinfo{journal}{Sci. Rep.}} \textbf{\bibinfo{volume}{8}},
  \bibinfo{pages}{3094} (\bibinfo{year}{2018}).

\bibitem{unicomb2019reentrant}
\bibinfo{author}{Unicomb, S.}, \bibinfo{author}{I{\~n}iguez, G.},
  \bibinfo{author}{Kert{\'e}sz, J.} \& \bibinfo{author}{Karsai, M.}
\newblock \bibinfo{title}{Reentrant phase transitions in threshold driven
  contagion on multiplex networks}.
\newblock \emph{\bibinfo{journal}{Phys. Rev. E}}
  \textbf{\bibinfo{volume}{100}}, \bibinfo{pages}{040301}
  (\bibinfo{year}{2019}).

\bibitem{karimi2013Athreshold}
\bibinfo{author}{Karimi, F.} \& \bibinfo{author}{Holme, P.}
\newblock \bibinfo{title}{Threshold model of cascades in empirical temporal
  networks}.
\newblock \emph{\bibinfo{journal}{Physica A}} \textbf{\bibinfo{volume}{392}},
  \bibinfo{pages}{3476--3483} (\bibinfo{year}{2013}).

\bibitem{karimi2013Btemporal}
\bibinfo{author}{Karimi, F.} \& \bibinfo{author}{Holme, P.}
\newblock \bibinfo{title}{A temporal network version of {W}atts's cascade
  model}.
\newblock In \bibinfo{editor}{Holme, P.} \& \bibinfo{editor}{Saramäki, J.}
  (eds.) \emph{\bibinfo{booktitle}{Temporal Networks}},
  \bibinfo{pages}{315--329} (\bibinfo{publisher}{Springer Berlin Heidelberg},
  \bibinfo{year}{2013}).

\bibitem{takaguchi2013bursty}
\bibinfo{author}{Takaguchi, T.}, \bibinfo{author}{Masuda, N.} \&
  \bibinfo{author}{Holme, P.}
\newblock \bibinfo{title}{Bursty communication patterns facilitate spreading in
  a threshold-based epidemic dynamics}.
\newblock \emph{\bibinfo{journal}{PloS One}} \textbf{\bibinfo{volume}{8}}
  (\bibinfo{year}{2013}).

\bibitem{backlund2014effects}
\bibinfo{author}{Backlund, V.-P.}, \bibinfo{author}{Saram{\"a}ki, J.} \&
  \bibinfo{author}{Pan, R.~K.}
\newblock \bibinfo{title}{Effects of temporal correlations on cascades:
  threshold models on temporal networks}.
\newblock \emph{\bibinfo{journal}{Phys. Rev. E}} \textbf{\bibinfo{volume}{89}},
  \bibinfo{pages}{062815} (\bibinfo{year}{2014}).

\bibitem{vazquez2006modeling}
\bibinfo{author}{V{\'a}zquez, A.} \emph{et~al.}
\newblock \bibinfo{title}{Modeling bursts and heavy tails in human dynamics}.
\newblock \emph{\bibinfo{journal}{Phys. Rev. E}} \textbf{\bibinfo{volume}{73}},
  \bibinfo{pages}{036127} (\bibinfo{year}{2006}).

\bibitem{whitt1982approximating}
\bibinfo{author}{Whitt, W.}
\newblock \bibinfo{title}{Approximating a point process by a renewal process,
  {I}: two basic methods}.
\newblock \emph{\bibinfo{journal}{Oper. Res.}} \textbf{\bibinfo{volume}{30}},
  \bibinfo{pages}{125--147} (\bibinfo{year}{1982}).

\bibitem{centola2007complex}
\bibinfo{author}{Centola, D.} \& \bibinfo{author}{Macy, M.}
\newblock \bibinfo{title}{Complex contagions and the weakness of long ties}.
\newblock \emph{\bibinfo{journal}{Am. J. Soc.}} \textbf{\bibinfo{volume}{113}},
  \bibinfo{pages}{702--734} (\bibinfo{year}{2007}).

\bibitem{valdano2015analytical}
\bibinfo{author}{Valdano, E.}, \bibinfo{author}{Ferreri, L.},
  \bibinfo{author}{Poletto, C.} \& \bibinfo{author}{Colizza, V.}
\newblock \bibinfo{title}{Analytical computation of the epidemic threshold on
  temporal networks}.
\newblock \emph{\bibinfo{journal}{Phys. Rev. X}} \textbf{\bibinfo{volume}{5}},
  \bibinfo{pages}{021005} (\bibinfo{year}{2015}).

\bibitem{yagan2012analysis}
\bibinfo{author}{Ya{\u{g}}an, O.} \& \bibinfo{author}{Gligor, V.}
\newblock \bibinfo{title}{Analysis of complex contagions in random multiplex
  networks}.
\newblock \emph{\bibinfo{journal}{Phys. Rev. E}} \textbf{\bibinfo{volume}{86}},
  \bibinfo{pages}{036103} (\bibinfo{year}{2012}).

\bibitem{gerstner2014neuronal}
\bibinfo{author}{Gerstner, W.}, \bibinfo{author}{Kistler, W.~M.},
  \bibinfo{author}{Naud, R.} \& \bibinfo{author}{Paninski, L.}
\newblock \emph{\bibinfo{title}{{N}euronal {D}ynamics: from {S}ingle {N}eurons
  to {N}etworks and {M}odels of {C}ognition}} (\bibinfo{publisher}{Cambridge
  University Press}, \bibinfo{year}{2014}).

\bibitem{iyer2013influence}
\bibinfo{author}{Iyer, R.}, \bibinfo{author}{Menon, V.},
  \bibinfo{author}{Buice, M.}, \bibinfo{author}{Koch, C.} \&
  \bibinfo{author}{Mihalas, S.}
\newblock \bibinfo{title}{The influence of synaptic weight distribution on
  neuronal population dynamics}.
\newblock \emph{\bibinfo{journal}{PLoS Comput. Biol.}}
  \textbf{\bibinfo{volume}{9}}, \bibinfo{pages}{e1003248}
  (\bibinfo{year}{2013}).

\bibitem{gleeson2011high}
\bibinfo{author}{Gleeson, J.~P.}
\newblock \bibinfo{title}{High-accuracy approximation of binary-state dynamics
  on networks}.
\newblock \emph{\bibinfo{journal}{Phys. Rev. Lett.}}
  \textbf{\bibinfo{volume}{107}}, \bibinfo{pages}{068701}
  (\bibinfo{year}{2011}).

\bibitem{jiang2013calling}
\bibinfo{author}{Jiang, Z.-Q.} \emph{et~al.}
\newblock \bibinfo{title}{Calling patterns in human communication dynamics}.
\newblock \emph{\bibinfo{journal}{Proc. Natl. Acad. Sci. U.S.A.}}
  \textbf{\bibinfo{volume}{110}}, \bibinfo{pages}{1600--1605}
  (\bibinfo{year}{2013}).

\bibitem{sorribes2011origin}
\bibinfo{author}{Sorribes, A.}, \bibinfo{author}{Armendariz, B.~G.},
  \bibinfo{author}{Lopez-Pigozzi, D.}, \bibinfo{author}{Murga, C.} \&
  \bibinfo{author}{de~Polavieja, G.~G.}
\newblock \bibinfo{title}{The origin of behavioral bursts in decision-making
  circuitry}.
\newblock \emph{\bibinfo{journal}{PLoS Comput. Biol.}}
  \textbf{\bibinfo{volume}{7}} (\bibinfo{year}{2011}).

\bibitem{saramaki2015exploring}
\bibinfo{author}{Saram{\"a}ki, J.} \& \bibinfo{author}{Holme, P.}
\newblock \bibinfo{title}{Exploring temporal networks with greedy walks}.
\newblock \emph{\bibinfo{journal}{Eur. Phys. J. B}}
  \textbf{\bibinfo{volume}{88}}, \bibinfo{pages}{334} (\bibinfo{year}{2015}).

\bibitem{eckmann2004entropy}
\bibinfo{author}{Eckmann, J.-P.}, \bibinfo{author}{Moses, E.} \&
  \bibinfo{author}{Sergi, D.}
\newblock \bibinfo{title}{Entropy of dialogues creates coherent structures in
  e-mail traffic}.
\newblock \emph{\bibinfo{journal}{Proc. Natl. Acad. Sci. U.S.A.}}
  \textbf{\bibinfo{volume}{101}}, \bibinfo{pages}{14333--14337}
  (\bibinfo{year}{2004}).

\bibitem{karimi2014structural}
\bibinfo{author}{Karimi, F.}, \bibinfo{author}{Ramenzoni, V.~C.} \&
  \bibinfo{author}{Holme, P.}
\newblock \bibinfo{title}{Structural differences between open and direct
  communication in an online community}.
\newblock \emph{\bibinfo{journal}{Physica A}} \textbf{\bibinfo{volume}{414}},
  \bibinfo{pages}{263--273} (\bibinfo{year}{2014}).

\bibitem{abate2006unified}
\bibinfo{author}{Abate, J.} \& \bibinfo{author}{Whitt, W.}
\newblock \bibinfo{title}{A unified framework for numerically inverting
  {L}aplace transforms}.
\newblock \emph{\bibinfo{journal}{INFORMS Journal on Computing}}
  \textbf{\bibinfo{volume}{18}}, \bibinfo{pages}{408--421}
  (\bibinfo{year}{2006}).

\end{thebibliography}
